\documentclass[aps,twocolumn,prd,nofootinbib,showpacs]{revtex4}

\usepackage{amsmath}
\usepackage{amssymb}
\usepackage{graphicx}

\def\({\left(} \def\){\right)}
\def\[{\left[} \def\]{\right]}
\def\dprime{\mathaccent"707D}
\def\vb{\bar v}
\def\bdi#1{\ensuremath{\boldsymbol{#1}}}
\def\bx{{\bdi x}}

\def\bX{{\bdi X}}

\def\bk{{\bdi k}}

\def\Xd{\dot X}
\def\Xdd{\ddot X}
\def\Xp{\acute X}
\def\Xpp{\dprime X}
\def\bXd{\dot\bX}
\def\bXp{\acute\bX}
\def\bXpp{\dprime{\bX}}

\def\del{\mbox{\boldmath $\nabla$}}

\def\tb{\bar{t}}

\newcommand\ba{\begin{array}}
\newcommand\ea{\end{array}}
\newcommand\ben{\begin{equation}}
\newcommand\een{\end{equation}}
\newcommand\bea{\begin{eqnarray}}
\newcommand\eea{\end{eqnarray}}

\newcommand{\ga}{\gamma}
\newcommand{\de}{\delta}
\newcommand{\ep}{\varepsilon}

\newcommand{\ka}{\kappa}
\newcommand{\la}{\lambda}

\newcommand{\si}{\sigma}
\newcommand{\ta}{\tau}

\newcommand{\Ga}{\Gamma}

\newcommand{\half}{\frac{1}{2}}

\newcommand{\vev}[1]{\langle#1 \rangle}
\newcommand{\lvev}[1]{\left\langle#1 \right\rangle}
\newcommand{\pa}{\partial}

\newcommand{\fd}{f_{10}}
\newcommand{\fnl}{f_{\mathrm{NL}}}
\newcommand{\deTovT}{\Theta}
\newcommand{\BT}{B}
\newcommand{\Area}{\mathcal{A}}
\newcommand{\fNL}{\fnl}
\newcommand{\fNLloc}{\fnl^{\mathrm{loc}}}
\newcommand{\fNLeq}{\fnl^{\mathrm{eq}}}

\newcommand{\sss}[1]{{\scriptscriptstyle{#1}}}
\newcommand{\vect}[1]{\boldsymbol{#1}}

\newcommand{\Gpc}{\textrm{Gpc}}
\newcommand{\zero}{{\sss{0}}}
\newcommand{\uCMB}{\mathrm{\sss{CMB}}}

\newcommand{\ures}{\mathrm{res}}
\newcommand{\ufov}{\mathrm{fov}}

\newcommand{\uc}{\mathrm{c}}

\newcommand{\ud}{\mathrm{d}}
\newcommand{\ui}{\mathrm{i}}

\newcommand{\ueq}{\mathrm{eq}}

\newcommand{\uiso}{\mathrm{iso}}
\newcommand{\uh}{\mathrm{h}}

\newcommand{\uL}{\mathrm{L}}

\newcommand{\Tcmb}{T_\uCMB}
\newcommand{\LCDM}{\Lambda\mathrm{CDM}}

\newcommand{\kperp}{k}

\newcommand{\unitn}{\vect{\hat{n}}}

\newcommand{\boxmpc}{L_\mathrm{sim}}
\newcommand{\corrini}{\ell_\uc}
\newcommand{\horizonini}{d_{\uh_\zero}}
\newcommand{\newton}{G}
\newcommand{\tension}{U}      
\newcommand{\GU}{\newton \tension}

\newcommand{\ang}{\theta}
\newcommand{\angres}{\ang_\ures}
\newcommand{\angfov}{\ang_\ufov}
\newcommand{\angsqz}{\ang}
\newcommand{\angcol}{\varphi}

\newcommand{\npixel}{n_\mathrm{pix}}

\newcommand{\FFTheta}[1]{\hat{\deTovT}_{{#1}}}
\newcommand{\window}[2]{W_{\negthinspace #1}\negthinspace\left(#2\right)}
\newcommand{\rbis}{b}
\newcommand{\calH}{\mathcal{H}}

\begin{document}
\title{The CMB temperature bispectrum induced by cosmic strings}
\author{Mark Hindmarsh}
\email{m.b.hindmarsh@sussex.ac.uk}
\affiliation{Department of Physics \& Astronomy, University of Sussex,
 Brighton, BN19QH, United Kingdom}
\author{Christophe Ringeval}
\email{christophe.ringeval@uclouvain.be}
\affiliation{Theoretical and Mathematical Physics Group, Centre for
 Particle Physics and Phenomenology, Louvain University, 2 Chemin du
 Cyclotron, 1348 Louvain-la-Neuve, Belgium}
\author{Teruaki Suyama}
\email{teruaki.suyama@uclouvain.be}
\affiliation{Theoretical and Mathematical Physics Group, Centre for
 Particle Physics and Phenomenology, Louvain University, 2 Chemin du
 Cyclotron, 1348 Louvain-la-Neuve, Belgium}
\date{\today}
\begin{abstract}
 The Cosmic Microwave Background (CMB) bispectrum of the temperature
 anisotropies induced by a network of cosmic strings is derived for
 small angular scales, under the assumption that the principal cause
 of temperature fluctuations is the Gott--Kaiser--Stebbins (GKS)
 effect. We provide analytical expressions for all isosceles
 triangle configurations in Fourier space. Their overall amplitude is
 amplified as the inverse cube of the angle and diverges for flat
 triangles. The isosceles configurations generically lead to a
 negative bispectrum with a power law decay $\ell^{-6}$  for large
  multipole $\ell$. However, collapsed triangles are found to be
 associated with a positive bispectrum whereas the squeezed triangles
 still exhibit negative values. We then compare our analytical
 estimates to a direct computation of the bispectrum from a set of
 $300$ statistically independent temperature maps obtained from
 Nambu--Goto cosmic string simulations in a
 Friedmann--Lema\^{\i}tre--Robertson--Walker (FLRW) universe. We find
 good agreement for the overall amplitude, the power law behaviour
 and angle dependency of the various triangle configurations. At
 $\ell \sim 500$ the cosmic string GKS effect contributes
 approximately the same equilateral CMB bispectrum amplitude as an
 inflationary model with $|\fNLloc| \simeq 10^3$, if the strings
 contribute about $10\%$ of the temperature power spectrum at
 $\ell=10$. Current bounds on $\fNL$ are not derived using 
 cosmic string bispectrum templates, and 
so our $\fNL$ estimate cannot be used to derive bounds on strings. 
However it does suggest that string bispectrum templates
 should be included in the search of CMB non-Gaussianities.
\end{abstract}
\maketitle
\section{Introduction}
\label{sec:intro}
Cosmic strings are line-like objects formed in the early universe
\cite{Hindmarsh:1994re,VilShe94,Sakellariadou:2006qs}.  They could be
solitons in field theories with spontaneously broken symmetries
\cite{Kibble:1976sj}, singular solutions with cylindrical symmetry in
supergravity theories \cite{Dabholkar:1990yf} or fundamental objects
in string theory \cite{Copeland:2003bj}.  They may form in thermal
phase transitions \cite{Kibble:1976sj}, at the end of hybrid inflation
\cite{Yokoyama:1989pa, Kofman:1994rk, Copeland:1994vg,
 Jeannerot:2003qv}, or by tachyon condensation at the end of brane
inflation when brane and anti-brane annihilate
\cite{Sarangi:2002yt,Dvali:2003zj}.  If cosmic strings are added to
the standard power-law $\Lambda$CDM model, the Cosmic Microwave
Background (CMB) data is fitted even better
\cite{Battye:2006pk,Bevis:2007gh} if the fraction of the temperature
power spectrum (at $\ell=10$) due to strings $\fd$ is about $0.1$.
There is therefore strong motivation to develop further tests for
strings in future CMB data, which provide the cleanest and best
understood cosmological string
signals~\cite{Gangui:2001fr,Senatore:2009gt,Smith:2009jr}. Calculations
of the polarisation B-mode \cite{Bevis:2007qz,Pogosian:2007gi} show
that a promising line of attack for the near future is to look for a
signal peaked between $\ell=600$ to $1000$. Simulations of Planck data
show that it will be sensitive down to $\fd \simeq 0.01$, and that
there is no danger of confusing strings with inflationary tensor
perturbations~\cite{Bevis:2007qz}.  Perhaps the most characteristic
signal in the CMB comes from the Gott-Kaiser-Stebbins (GKS)
effect~\cite{Gott:1985,Kaiser:1984}, which is due to the gravitational
lensing of photons passing near a moving string. This produces a
discontinuity in the apparent temperature approximately proportional
to the transverse velocity of the string $v$ and the string tension
$\tension$:
\begin{equation}
\de T \sim 8\pi
(G\tension)v \Tcmb,
\end{equation}
where $G$ is Newton's constant. Given that strings move with a
mean square (RMS) velocity of between $0.25$ and
$0.36$~\cite{Hindmarsh:2008dw,Martins:2006} we would expect to see
discontinuities $\de T$ of up to about 1 $\mu$K.
Recent calculations of the integrated Sachs-Wolf CMB signal at small
angular scales have shown that the angular power spectrum it produces
decreases slowly, approximately $\ell^{-0.9}$ at high multipole moment
$\ell$ \cite{Fraisse:2007nu}.  If $\fd \simeq 0.1$ strings should
dominate the adiabatic temperature power spectrum for $\ell \gtrsim
3000$, and remain above the thermal Sunyaev-Zel'dovich effect.  This
fraction corresponds to $G\tension \simeq 0.7\times 10^{-6}$ for
strings in the Abelian Higgs model \cite{Bevis:2007gh}.  The slow
power-law decrease in the power spectrum is quite close to a
prediction of the small scale power spectrum using the string
correlation functions in a Gaussian approximation
\cite{Hindmarsh:1993pu}, which gave a temperature anisotropy power
going as $\ell^{-1}$.  The non-Gaussian nature of the maps exhibited
in \cite{Fraisse:2007nu}, as for example exhibited by the skewness of
the distribution of the 1-point function, immediately motivates an
attempt to generalise the calculation to higher-order correlators, and
in particular the 3-point function or bispectrum, which will become
increasingly well characterised by future CMB
data~\cite{Spergel:1999xn, Gangui:2000gf, Komatsu:2008hk,
 Smith:2009jr}.  This paper reports on the results of this
calculation.  It is found that the bispectrum  for an isosceles
arrangement of wavevectors $\bk$ in Fourier space is generically
negative and decreases as $|\bk|^{-6}$. It is proportional to
$(G\tension)^3$ and hence to the power spectrum raised to the $3/2$
power.  The scale of the non-linearity parameter $\fnl$, defined by
dividing the bispectrum by the square of the power spectrum, is
therefore potentially large, going as $(G\tension)^{-1}$.

\section{Analytical bispectrum}
\label{sect:analytical}
\subsection{Gott-Kaiser-Stebbins effect}
We work in the flat sky approximation, and define transverse
coordinates $\bx$ measured in radians. The wave number $\bk$ is
related to the multipole moment $\ell$ by~\cite{White:1999,
 LoVerde:2008re}
\begin{equation}
 k^2 \simeq \ell(\ell + 1).
\end{equation}
The anisotropy power $\ell(\ell+1)C_\ell$ is then approximately equal
to $k^2|\de T_\bk|^2$, where $\de T_\bk$ is the Fourier transform of
the temperature fluctuation,
\begin{equation}
 \de T_{\bk} = \int \ud \bx {\de T}
 e^{i\bk\cdot\bx}.
\end{equation}
We will also define $\deTovT(\bx) = \de T(\bx)/\Tcmb$.
String spacetime coordinates will be denoted $X^\mu(\tau,\si)$, where
$\tau$ and $\sigma$ are timelike and spacelike worldsheet coordinates
respectively.  In the temporal gauge the $X^0 = \tau$ (where the
worldsheet time is identified with the background time coordinate) the
(corrected) GKS formula is \cite{Hindmarsh:1993pu}
\begin{equation}
\del^2 \deTovT = -8\pi G\tension \int \ud\si \left[ \bXd - {(\Xp \cdot \hat p)
 \over (\Xd \cdot \hat p)} \bXp \right] \cdot \del \de^{(2)}(\bx - \bX),
\label{eTemGau}
\end{equation}
where $\hat p^\mu = p^\mu/E$, and worldsheet variables are evaluated
at the retarded time $t_{\rm r} = t+z - X^3(\si,t_{\rm r})$, when the
CMB photons (taken to be moving in the $-z$ direction) pass the
string.
The expression is greatly simplified in the light cone gauge,  
\begin{equation}
X^+(\si,\ta) = \ta,
\label{eLCG} 
\end{equation}
where $X^\pm = X^0 \pm X^3$.  
The time parameter $\ta$ then labels the intersections with a set of
null hyperplanes with the worldsheet: the photon geodesics $Z^\mu =
x^\mu + \la p^\mu$ form just such a set.  Then we find
\begin{equation}
\del^2 \deTovT = -8\pi G\tension \int \ud\si \bXd\cdot \del \de^{(2)}(\bx - \bX),
\label{eTemFlu}
\end{equation}
where worldsheet quantities are now evaluated at $\ta=x^+=t+z$.
In Fourier space the equation becomes 
\begin{equation}
-k^2\deTovT_{\bk} = i\epsilon k_A   \int \ud\si \Xd^A(\si)
e^{i\bk\cdot\bX(\si)},
\label{eFT}
\end{equation} where we have defined
\begin{equation}
 \epsilon = 8\pi G\tension,
\end{equation}
and $A = 1,2$ with implicit summation on repeated indices. It is now
clear that the power spectrum, bispectrum, and higher order
correlators can be evaluated in terms of correlation functions of the
string network, as projected onto our backward light cone.  In the
next section we will introduce the relevant correlation functions and
discuss their important features.
\subsection{String correlation functions}
\begin{figure*}
\begin{center}
\includegraphics[width=0.32\textwidth]{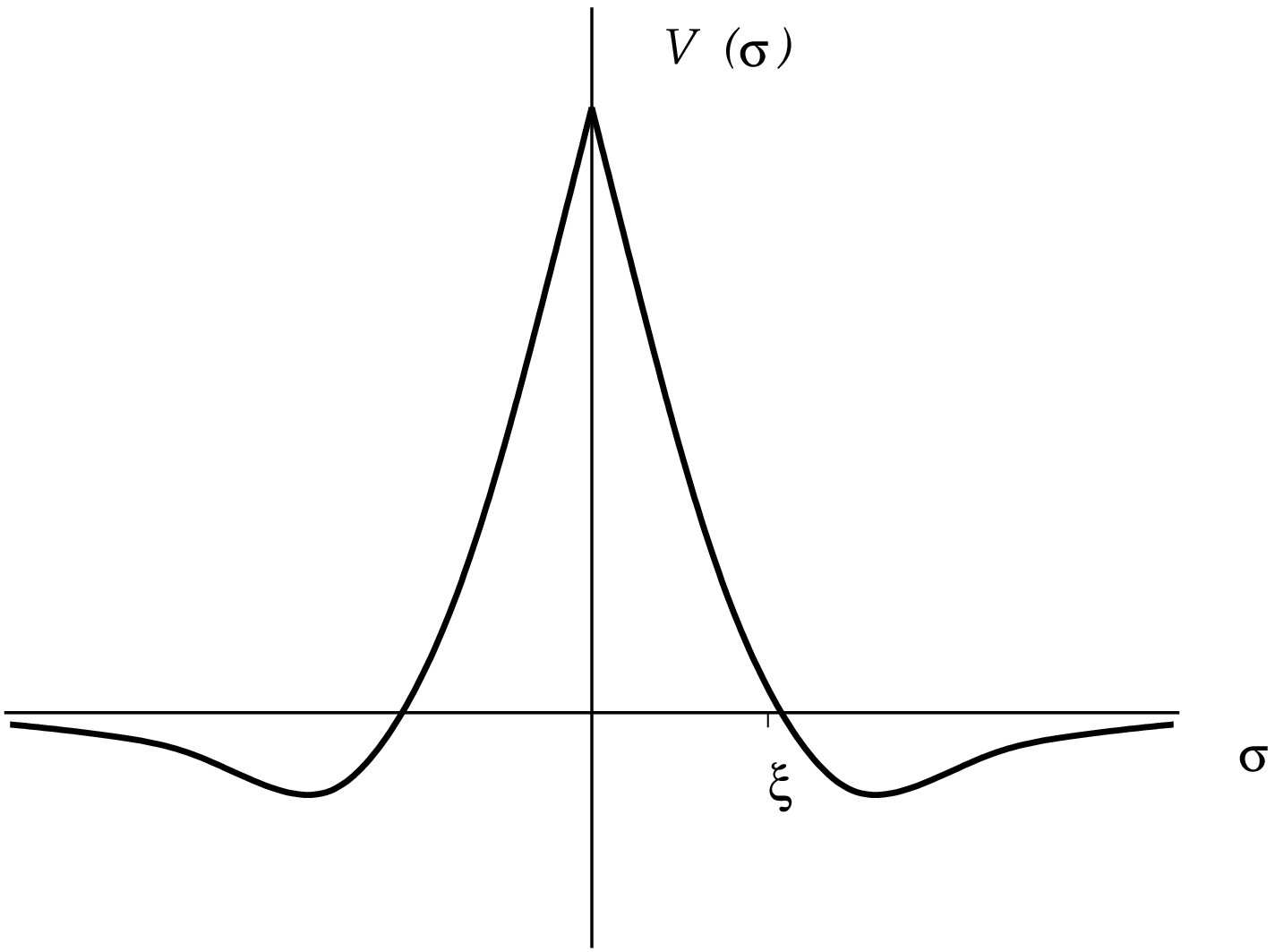}
\includegraphics[width=0.32\textwidth]{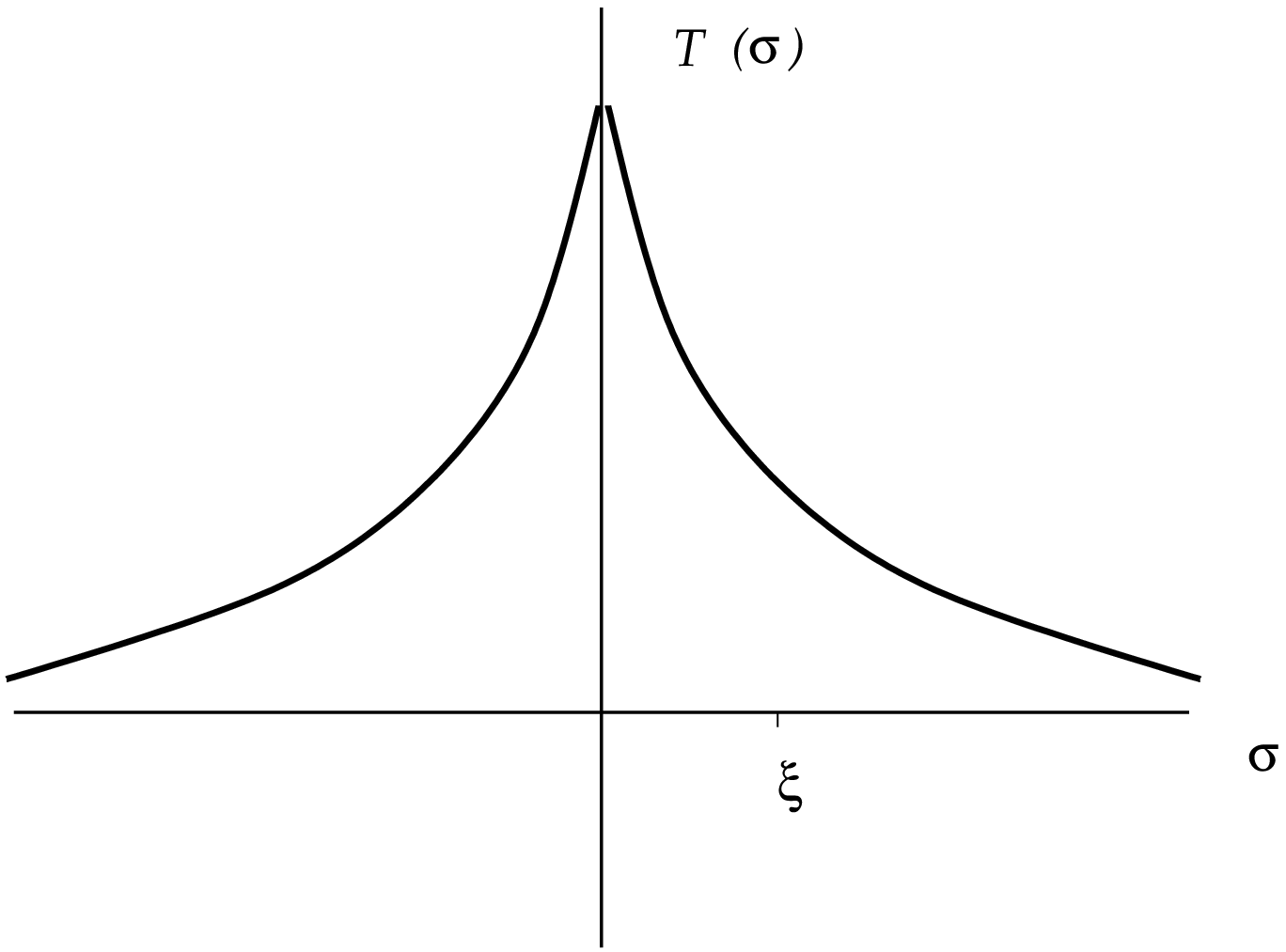}
\includegraphics[width=0.32\textwidth]{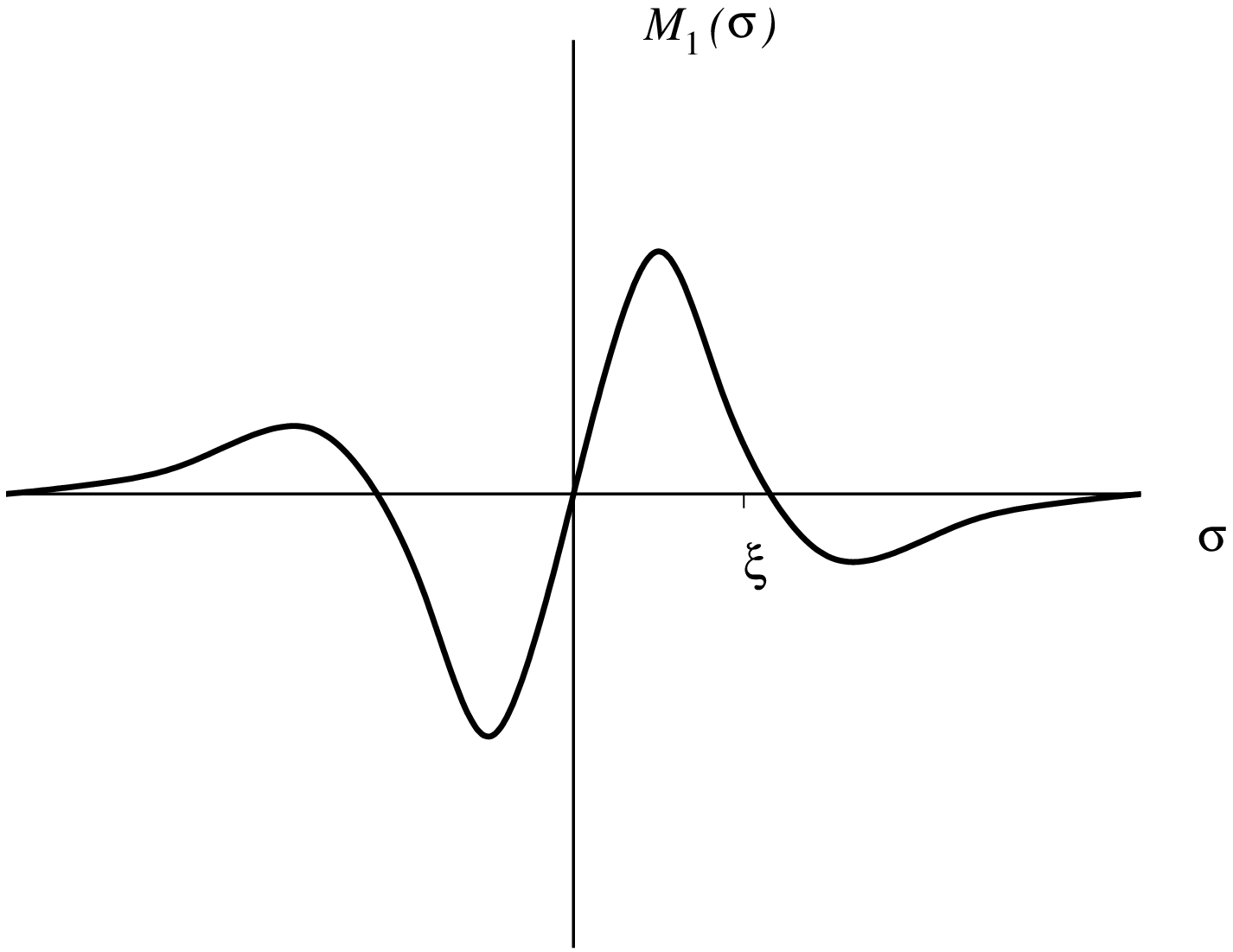}
\caption{Sketches of string correlation functions $V(\sigma)$
 (velocity-velocity), $T(\sigma)$ (tangent-tangent) and $M_1(\sigma)$
 (velocity-tangent), as a function of the string worldsheet spacelike
 separation $\sigma$, defined in Eqs.~(\ref{e:VelCor}) to
 (\ref{e:MCor})}.
\label{f:CorSketch}
\end{center}
\end{figure*}

We denote the transverse coordinates of the string, $X^A(\si)$.  The
basic two point functions are
\begin{equation}
\begin{aligned}
\vev{\Xd^A(\si)\Xd^B(\si')},\quad \vev{\Xd^A(\si)\Xp^B(\si')},\quad
\vev{\Xp^A(\si)\Xp^B(\si')},
\end{aligned}
\end{equation}
where the angle brackets denote an average over an ensemble of
strings.  The starting assumption is that the string ensemble is
well-approximated by a Gaussian process: that is, all correlators can
be calculated in terms of the two point functions.
We now make some assumptions about the ensemble: (\romannumeral 1)
rotation, reflection and translation invariance of the transverse
coordinates; and (\romannumeral 2) worldsheet reflection and
translation invariance.  Then there are three independent correlation
functions: 
\begin{eqnarray}
\vev{\Xd^A(\si)\Xd^B(\si')} &=& \half\de^{AB}V(\si-\si'), \label{e:VelCor}\\
\vev{\Xp^A(\si)\Xp^B(\si')} &=& \half \de^{AB} T(\si-\si'), \label{e:TanCor}\\
\vev{\Xp^A(\si)\Xd^B(\si')} &=&
\half\de^{AB}M_1(\si-\si'). \label{e:MCor}
\end{eqnarray}
A fourth, 
\begin{equation}
M_2(\si) = \vev{\Xd^A(\si)\Xp^B(0)} \epsilon^{AB},
\end{equation}
vanishes because of the symmetry $X^1 \leftrightarrow X^2$ (this point
was overlooked in Ref.~\cite{Hindmarsh:1993pu}). The functions $V$ and
$T$ are symmetric in their argument, while $M_1$ is antisymmetric.
The forms of the correlators are sketched in Fig.~\ref{f:CorSketch} (see also \cite{Martins:2006}).
For later convenience two other correlators will be defined:
\begin{align}
 \Ga(\si-\si') & \equiv \lvev{\left[ \vect{X}(\si)-\vect{X}(\si')\right]^2} \\ & =
 \int_{\si'}^\si \ud\si_1\int_{\si'}^\si \ud\si_2 T(\si_1-\si_2)
 , \\
 \Pi(\si-\si') & \equiv
 \lvev{\left[\vect{X}(\si)-\vect{X}(\si'))\right] \cdot \vect{\Xd}(\si')} \\ &=
 \int_{\si'}^\si \ud\si_1M_1(\si_1-\si') , \label{e:MixCor}
\end{align}
There are integral constraints arising from the fact that the average velocity of the 
strings vanishes on large scale:
\begin{equation}
\int \ud\si V(\si) \to 0, \quad \int \ud\si M_1(\si) \to 0.
\end{equation}
The important asymptotic forms are:
\begin{eqnarray}
 V(\si) & \to & \left\{ \ba{cl}  \vb^2 & \si \to 0 \\ 0 & \si \to \infty  \ea \right. ,\\
 \Ga(\si) & \to & \left\{ \ba{cl}  \tb^2\si^2 & \si \to 0 \\
   \hat\xi\si & \si \to \infty  \ea \right. , \label{e:AsymGamma}\\
 \Pi(\si) & \to & \left\{ \ba{cl}  \half \frac{c_0}{\hat\xi}\si^2 &
   \si \to 0 \\
   0 & \si \to \infty  \ea \right. ,
\end{eqnarray}
where we have defined four parameters
\begin{eqnarray}
\hat\xi &=& \Ga'(\infty), \label{e:xi}\\
\vb^2 &=& \lvev{\bXd^2},\label{e:vbar}\\
\tb^2 &=& \lvev{\bXp^2}, \label{e:tbar}\\
c_0 &=& \hat\xi\lvev{\bXpp\cdot\bXd}. \label{e:c0}
\end{eqnarray}
The correlation length $\hat\xi$ is the projected correlation length
on the backward lightcone, $\bar t^2$ is the mean square projected
tangent vector (of order unity), $\bar v^2$ is the mean square
projected velocity (again of order unity), and $c_0$ is the
correlation between the projected velocity and curvature. 

\subsection{Light cone gauge equations}
\label{ss:lcg}
In Minkowski space, the Nambu--Goto action leads to the equations of
motion and constraints
\begin{equation}
\Xdd^\mu - \Xpp^\mu = 0, \quad \Xd^2 + \Xp^2 = 0, \quad \Xd\cdot\Xp = 0.
\end{equation}
The light-cone gauge consists of choosing $\tau = X^+ =
X^0+X^3$. Hence
\begin{equation}
\Xd^- = \Xd_A\Xd^A + \Xp_A\Xp^A, \quad \Xp^- = 2\Xd_A\Xp^A.
\end{equation}
The equations of motion in a FLRW background follow from the
Nambu--Goto action
\begin{eqnarray}
S &=& -\tension \int \ud\ta \ud\si \sqrt{-\ga} \nonumber\\ 
&=& - \tension \int \ud\ta \ud\si a^2(X^0) \sqrt{-\Xd^2\Xp^2 + (\Xd\cdot\Xp)^2},
\end{eqnarray}
where $\gamma$ is the determinant of the induced metric along the
string worldsheet, and $a(X^0)$ is the scale factor. With the standard
gauge choice $\Xd\cdot\Xp = 0$, we find 
\begin{equation}
\begin{aligned}
 \Xdd^\mu  & + \( \frac{\dot\ep}{\ep}  + 2 \frac{\dot a}{a}\) \Xd^\mu  -
 \frac{1}{\ep} \frac{\pa}{\pa\si}\(\frac{1}{\ep} \frac{\pa
   X^\mu}{\pa\si} \) 
 \\ & - 2 \frac{1}{a} \frac{\ud a}{\ud X^0}\frac{1}{\ep}\frac{\pa
   X^0}{\pa\si}\frac{1}{\ep} \frac{\pa X^\mu}{\pa\si} + 2 \de^\mu_0 \frac{1}{a}
 \frac{\ud a}{\ud X^0}\Xd^2 = 0,
\end{aligned}
\end{equation}
where $\ep = \sqrt{-\Xp^2/\Xd^2}$.
The light-cone gauge choice $X^+=\ta$ produces as the equation of
motion for $X^+$ 
\begin{equation}
\frac{\dot\ep}{\ep} + 2 \calH(\Xd^0 + \Xd^2) = 0,
\end{equation}
where $\calH=\ud{\ln a}/\ud X^0$ is the conformal
Hubble parameter.  The equation for the transverse components is 
\begin{equation}
\begin{aligned}
 \ddot\bX + 2 \calH\frac{1}{\ep^2}\( \bXp^2\) \bXd & -
 \frac{1}{\ep} \frac{\pa}{\pa\si}\(\frac{1}{\ep} \frac{\pa
   \bX}{\pa\si} \)
 \\
 &- 2 \calH \frac{1}{\ep^2}(\bXd\cdot\bXp) \bXp = 0.
\end{aligned}
\end{equation}
In a FLRW background, assuming that $\vev{\bXd^2}$ is constant, and
neglecting higher-order correlations between $\calH$, $\bXd$ and $\bXp$, we find
\begin{equation}
\begin{aligned}
 \lvev{\frac{\pa^2\bX}{\pa s^2}\cdot\bXd} & = 2 \bar\calH \lvev{
     \left(\frac{\pa \bX}{\pa s}\right)^2\bXd^2} \\ & - 2 \bar{\calH}
   \lvev{\left(\bXd\cdot\frac{\pa\bX}{\pa s}\right)^2} ,
\label{e:XppXdCorr}
\end{aligned}
\end{equation}
where we have defined $\ud s = \ep \ud\si$, and where $\bar \calH$ is
the conformal Hubble parameter averaged over the string ensemble at
fixed $\tau=X^+$, which will select the value where there is most
string, i.e. at decoupling.

If one we assume that the string ensemble is approximately
Gaussian in $\bXd$ and $\bXp/\ep$, it is not hard to show that the
right hand side reduces to
\begin{equation}
\begin{aligned}
\lvev{\dfrac{\partial^2 \bX}{\partial s^2} \cdot \bXd} & = \bar{\calH}
\left( \lvev{\bXd^2} \lvev{\bXp^2} - \lvev{\bXd \cdot \bXp}^2 \right).
\end{aligned}
\end{equation}
The last term vanishes, so the cross correlator simplifies to
$\bar{\calH} \vb^2 \tb^2$ which is positive. From Eq.~(\ref{e:c0}),
one gets that in a FLRW background, a string network should exhibit
$c_0 > 0$. We see also that it vanishes in Minkowski space, which can
be viewed as a consequence of time reversal invariance. As shown in
the next sections, the temperature bispectrum vanishes when the
correlator $\Pi$ does, and so the generation of a bispectrum by
strings simply requires the breaking of time reversal invariance, as
it is in a FLRW background\footnote{We thank B.~Wandelt for explaining
 this point to us.}.

\subsection{Temperature power spectrum at small angular scales}
The power spectrum at small angular scales, where the GKS
effect \cite{Gott:1985,Kaiser:1984iv} is held to be dominant, was calculated in
Ref.\ \cite{Hindmarsh:1993pu}, and the calculation is recapped here
for completeness.
The Fourier transform of the temperature fluctuation given in
Eq.~(\ref{eTemFlu}) reads
\begin{equation}
\deTovT_{\bk} = \int \ud^2 x \frac{\de T}{T} e^{i\bk\cdot\bx},
\end{equation}
and hence is given by Eq.~(\ref{eFT}). The power spectrum is defined by
\begin{equation}
\lvev{\deTovT_{\bk_1}\deTovT_{\bk_2}} = P(k_1)(2\pi)^2\de(\bk_1+\bk_2).
\end{equation}
With our conventions we need a box of formal area $\Area=(2\pi)^2\de(0)$
to express the power spectrum
\begin{equation}
\begin{aligned}
P(k) 
& = \epsilon^2 {k_Ak_B\over A k^4} \int \ud\si \ud\si' 
\left\langle
\Xd^A(\si)\Xd^B(\si') \right. \\ & \left. 
\times e^{i\bk\cdot [\bX(\si)-\bX(\si')]}\right\rangle.
\end{aligned}
\end{equation}
With our assumptions about the string correlation functions, the
ensemble average can be reduced to
\begin{equation}
\begin{aligned}
 P(k) & = \half\ep^2 {1\over \Area k^2}\int \ud\si \ud\si'\left[
   V(\si-\si') \phantom{\half}\right. \\ & + \left. \half k^2 \Pi^2(\si-\si')
 \right] e^{-k^2\Ga(\si-\si')/4}.
\end{aligned}
\label{ePowSpe}
\end{equation}
We now derive the asymptotic behaviour of $P(k)$ as $(k\xi)$ gets
large.  The contribution to the power spectrum from the mixed
correlator $M_1$ can be shown to be subdominant at high $k$
\cite{Hindmarsh:1993pu}, and so we need examine only the first term in
the power spectrum of Eq.~(\ref{ePowSpe}),
\begin{equation}
 P(k) = \ep^2 {1\over 4\Area k^2}  \int \ud\si_+ \ud\si_- V(\si_-)
 e^{-k^2\Ga(\si_-)/4},
\end{equation}
where $\si_\pm = \si\pm\si'$.  For $k\hat\xi \gg 1$, we find
\begin{equation}
 k^2P(k) \simeq  \ep^2  \sqrt{\pi} \frac{L\hat\xi}{\Area} \frac{\vb^2}{\tb}
 {1\over (k\hat\xi)}\, ,
\end{equation}
where $L$ is the total transverse light-cone gauge length of string in
the box of area $\Area$.

The power spectrum given in Ref.~\cite{Bouchet:1988hh} was consistent
with $k^{-1}$, and the amplitude was surprisingly close for such a
crude estimate. Fraisse et al.~\cite{Fraisse:2007nu} have a slightly
different small-scale angular power spectrum: $ k^2P(k) \sim k^{-p}$
with $p \simeq -0.89$.  This can be explained if $\Ga(\si) \sim
\si^{2/p}$ on the relevant scales. This correlation function controls
how far on average one moves in the transverse coordinates as one
moves along the string: $p=1$ would correspond to straight lines,
while $p=2$ to a Brownian random walk.  A power less than one is
suggestive of a cloud of zero-dimensional objects along the string
worldsheet which may be the signature of small loop production.

\subsection{Temperature bispectrum from strings at small angular
 scales}

In the flat sky approximation the three point temperature correlation
function or bispectrum is defined as
\begin{equation}
\vev{\deTovT_{\bk_1}\deTovT_{\bk_2}\deTovT_{\bk_3}} =
\BT(\bk_1,\bk_2,\bk_3)(2\pi)^2\de(\bk_1+\bk_2+\bk_3).
\end{equation}
We again need to normalise by a formal area factor $\Area =
(2\pi)^2\de(0)$ to obtain an expression in terms of a string
expectation value
\begin{equation}
\begin{aligned}
\BT(\bk_1,\bk_2,\bk_3) &= i\epsilon^3\frac{1}{\Area} \delta_{A
 \bar{A}}\delta_{B \bar{B}} \delta_{C \bar{C} }\frac{k^{\bar{A}}_{1}
 k^{\bar{B}}_{2} k^{\bar{C}}_{3}}{k_1^2k_2^2k_3^2} \\ & \times \int
\ud\si_1\ud\si_2\ud\si_3 \lvev{\Xd^A_1\Xd^B_2\Xd^C_3 e^{i \delta^{ab}
   \bk_a \cdot \bX_b}},
\end{aligned}
\end{equation}
with $\Xd^A_a = \Xd^A(\si_a)$, $a,b \in \{1,2,3\}$, and
$\bk_1+\bk_2+\bk_3=0$. With the Gaussian assumption, the ensemble
average of the string observables is lengthy but straightforward.
Defining
\begin{equation}
C^{ABC} \equiv \Xd^A_1\Xd^B_2\Xd^C_3,\quad D \equiv \delta^{ab} \bk_a \cdot
\bX_b,
\end{equation}
we have
\begin{equation}
\lvev{C^{ABC} e^{iD}} = i\lvev{C^{ABC}D} e^{-\half \vev{D^2}},
\end{equation}
and hence
\begin{equation}
\begin{aligned}
\BT(\bk_1,\bk_2,\bk_3) &= -\epsilon ^3\frac{1}{\Area}
\dfrac{k_{1_A}k_{2_B}k_{3_C}} {k_1^2k_2^2k_3^2} \\ & \times \int
\ud\si_1 \ud \si_2 \ud\si_3 \lvev{C^{ABC}D} e^{-\half \vev{D^2}}.
\label{e:CDint}
\end{aligned}
\end{equation}
We first evaluate the correlator
\begin{equation}
\begin{aligned}
&\lvev{C^{ABC}D} = \lvev{\Xd^A_1\Xd^B_2\Xd^C_3 \, \bk^a \cdot \bX_a}
 \\ & =  \vev{\Xd^A_1\Xd^B_2\ } \vev{\Xd^C_3 \bk^a \cdot \bX_a } + 
\vev{\Xd^C_3\Xd^A_1\ } \vev{\Xd^B_2 \bk^a \cdot \bX_a } \\ &+ 
\vev{\Xd^B_2\Xd^C_3\ } \vev{\Xd^A_1 \bk^a \cdot \bX_a }.
\end{aligned}
\end{equation}
The velocity correlators have already been given in
Eq.~(\ref{e:VelCor}).  The mixed correlators are most easily evaluated
using variations on the theme
\begin{equation}
\begin{aligned}
\bk^a \cdot \bX_a & = \bk_1\cdot(\bX_1 - \bX_3) + \bk_2\cdot(\bX_2 -
\bX_3) \\ 
& = \bk_1 \cdot \bX_{13} + \bk_2 \cdot \bX_{23},
\end{aligned}
\end{equation}
where $\bX_{ab} \equiv \bX_a - \bX_b$. For example,
\begin{equation}
\begin{aligned}
\lvev{\Xd^A_1 \bk^a \cdot \bX_a } &= k_2^{D} \lvev{\Xd^A_1 X^D_{21} } +
k_3^D \lvev{\Xd^A_1 X^D_{31} } .
\end{aligned}
\end{equation}
Using the definition of the correlator $\Pi(\si)$ in
Eq.~(\ref{e:MixCor}),
\begin{eqnarray}
\lvev{\Xd^A_1 \bk^a \cdot \bX_a }  &=&  
\half k^A_2\Pi( \si_{21}) + \half k^A_3 \Pi( \si_{31}),
\end{eqnarray}
where $\si_{ab} = \si_a - \si_b$.
Substituting for the velocity correlators
\begin{equation}
\begin{aligned}
 \lvev{C^{ABC}D} &=  \frac14 \de^{AB} \[ k^C_1\Pi( \si_{13}) + k^C_2
 \Pi( \si_{23}) \] V(\si_{12}) \\ &+ \frac14 \de^{CA} \[ k^B_1\Pi(
 \si_{12}) + k^B_3 \Pi( \si_{32}) \] V(\si_{31}) \\ &+ \frac14
 \de^{BC} \[ k^A_2\Pi( \si_{21}) + k^A_3 \Pi( \si_{31}) \]
 V(\si_{23}). 
\end{aligned}
\end{equation}
Now we evaluate
\begin{equation}
\begin{aligned}
\lvev{D^2} & = \lvev{\left(\bk^a \cdot \bX_a\right)^2}
= \lvev{\( \bk_1\cdot \bX_{13}  +  \bk_2\cdot \bX_{23}  \)^2} \\
& =  \half k_1^2 \Ga(\si_{13}) + \half k_2^2 \Ga(\si_{23})  + \bk_1\cdot \bk_2 
\lvev{\bX_{13} \cdot \bX_{23} }.
\end{aligned}
\end{equation}
It can be shown that 
\begin{equation}
\lvev{\bX_{13} \cdot \bX_{23} } = \half \left[ \Ga(\si_{13}) +
 \Ga(\si_{23}) - \Ga(\si_{12}) \right],
\end{equation}
hence 
\begin{equation}
\begin{aligned}
\lvev{D^2} = -\half [\bk_1\cdot \bk_3 \Ga(\si_{13}) & +
 \bk_2\cdot\bk_3 \Ga(\si_{23})  \\ &  + \bk_1\cdot \bk_2
 \Ga(\si_{12})].
\end{aligned}
\end{equation}
Note that this expression is symmetric under the interchange of any
pair $\{\si_a,\si_b\}$ and $\{\bk_a,\bk_b\}$. Contracting
Eq.~(\ref{e:CDint}) with $k_1^Ak_2^Bk_3^C$, and defining
\begin{equation}
\ka_{ab} \equiv -\bk_a\cdot\bk_b,
\end{equation}
we have
\begin{equation}
\begin{aligned}
& \BT(\bk_1,\bk_2,\bk_3)  = - \epsilon^3 \frac{1}{\Area}
\dfrac{1}{4 k_1^2k_2^2k_3^2}\int \ud\si_1 \ud\si_2 \ud\si_3  \\ 
\times \bigg\{&\ka_{12} \[\ka_{13}\Pi(
\si_{13})  +  \ka_{23} \Pi( \si_{23}) \] V(\si_{12})
\\ + &\ka_{13} \[ \ka_{12}\Pi( \si_{12}) +
\ka_{23} \Pi( \si_{32}) \] V(\si_{31})  \\  +
& \ka_{23} \[ \ka_{12}\Pi( \si_{21})  + \ka_{13} \Pi(
\si_{31}) \] V(\si_{23})  \bigg\} \\ &  \times
\exp\left\{- \frac14 \[ \ka_{13} \Ga(\si_{13}) + \ka_{23}
\Ga(\si_{23}) + \ka_{12} \Ga(\si_{12}) \] \right\}.
\end{aligned}
\end{equation}
We perform the integrations over the string coordinates $\si_a$ in the
Appendix~\ref{sec:strginteg}, where it is found
\begin{equation}
\begin{aligned}
\BT(\bk_1,\bk_2,\bk_3) & = -\epsilon^3 \pi
c_0\frac{\vb^2}{\tb^4}\frac{L\hat\xi}{\Area}\frac{1}{\hat\xi^2}
\frac{1}{k_1^2k_2^2k_3^2} \\ & \times \left[ \frac{k_1^4
 \ka_{23} + k_2^4\ka_{31} + k_3^4\ka_{12}}{\(\ka_{23}\ka_{31}
 +\ka_{12}\ka_{31}+ \ka_{12}\ka_{23} \)^{3/2}} \right].
\end{aligned}
\label{e:Beqn}
\end{equation}

This is our primary expression for the bispectrum induced by the GKS
effect in cosmic strings.  It is proportional to $(G\tension)^3$, and
therefore goes as the $3/2$ power of the power spectrum. The factor
$L\hat\xi/\Area$ is a geometrical factor of order unity, as the
projected string length per unit area is of order the projected
correlation length $\hat\xi$ (unless there are a large number string
networks which do not interact with each other).  The factor in curly
brackets is geometrical in Fourier space, depending on the relative
lengths and angles of the $\bk_a$.  The overall dependence on angular
quantities is therefore $\hat\xi^{-2}k^{-6}$.  The dependence on
string correlators appears through $\bar t$, $\bar v^2$, and $c_0$
defined in Eqs.~(\ref{e:xi}) to (\ref{e:c0}). This last is an
interesting quantity as it is not time-reversal invariant: hence the
bispectrum could vanish only if the string network is time
symmetric. In an expanding universe, the existence of an asymmetry is
ensured by decay of the string network. It was argued at the end of
Section \ref{ss:lcg} that $c_0$ should be positive.

\subsection{Symmetrical triangle configurations}

\label{sec:triconfig}
Current analysis of the CMB temperature bispectrum focus essentially
on a particular local model of primordial non-Gaussianity as well as
specific configurations of the primordial bispectrum wavenumbers. From
Eq.~(\ref{e:Beqn}), we can easily estimate the string induced
bispectrum for various symmetrical triangle configurations in Fourier
space. Let us stress that Eq.~(\ref{e:Beqn}) is the CMB temperature
bispectrum as opposed to a primordial bispectrum which is the three
point-functions of the primordial Newtonian potential, usually of
inflationary origin. The latter has still to be evolved through the
CMB transfer functions. As a result, even for identical wavenumber
configurations in Fourier space, the resulting temperature bispectra
are not the same.

\subsubsection{Isosceles triangle}
We now consider isosceles triangle configurations in Fourier space such
that
\begin{equation}
\left|\bk_1\right| = \left|\bk_2\right|=k,\quad \left|\bk_3\right| = 2
k \sin \dfrac{\theta}{2},
\end{equation}
where $\theta$ denotes the angle between the wavevectors $\bk_1$ and
$\bk_2$. The cross scalar products simplify to
\begin{equation}
\kappa_{12} = k^2 \cos \theta, \quad \kappa_{23} = \kappa_{31} = 2 k^2 \sin^2(\theta/2),
\end{equation}
and the isosceles bispectrum reads
\begin{equation}
\label{eq:biso}
\BT_{\uiso}(k,\theta) = -\epsilon^3 \pi
c_0\frac{\vb^2}{\tb^4}\frac{L\hat\xi}{\Area} \dfrac{1}{\hat{\xi}^2 k^6}\dfrac{1 + 4 \cos
 \theta \sin^2(\theta/2)}{\sin^3\theta}\,.
\end{equation}
\begin{figure}
\begin{center}
\includegraphics[width=8.cm]{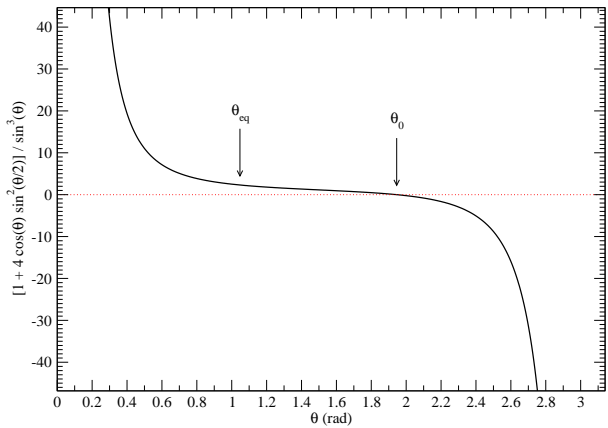}
\caption{Angular dependency of the isosceles bispectrum as a function
 of the angle $\theta$ in between the wavevectors $\bk_1$ and
 $\bk_2$. The particular values $\theta_\ueq = \pi/3$ corresponding
 to the equilateral configuration as $\theta_0$ at which the
 bispectrum vanishes are represented. Notice the divergences for flat
 triangle configurations at $\theta\to0$ (squeezed) and
 $\theta\to\pi$ (collapsed).}
\label{fig:biso}
\end{center}
\end{figure}

Notice that for $\theta=\pi/3$, we obtain the peculiar case of an
equilateral triangle [see Eq.~(\ref{eq:beq})]. In Fig.~\ref{fig:biso},
we have plotted the angle dependency of the isosceles bispectrum. Such
a configuration diverges in the two flat triangle limits for which
either $\theta \to 0$ or $\theta \to \pi$. Both of these
configurations are therefore better suited than the equilateral one to
characterize the strings and are discussed in the next
sections. Notice also the change of sign which occurs for the angle
\begin{equation}
\theta_0 = 2 \arccos\dfrac{\sqrt{3-\sqrt{3}}}{2}\,.
\end{equation}
For $\theta<\theta_0$, one has $\BT_{\uiso}<0$ (as for example in the
equilateral configuration), whereas for $\theta>\theta_0$ the
bispectrum $\BT_{\uiso} > 0$.

\subsubsection{Squeezed triangles}

This is the case of one of the sides of the triangle vanishing $k_3\to
k\theta$, with the opposite angle $\theta\to 0 $. The angular factor
in Eq.~(\ref{eq:biso}) simplifies to $1/\theta^3$ and
\begin{equation}
 \BT_{kk\theta} \equiv \lim_{\theta\to 0} \BT_{\uiso}(k,\theta)
 \underset{\theta \to 0}{\sim}
 -\epsilon^3 {\pi}
 c_0\frac{\vb^2}{\tb^4}\frac{L\hat\xi}{\Area}
 \frac{1}{\hat\xi^2k^6} \dfrac{1}{\theta^3}\,.
\end{equation}
The bispectrum is therefore negative and appears to diverge in the
squeezed limit. In practice, it is not possible to observe such a
divergence as the maximum value of the wave vector is bounded by the
detector resolution, and the minimum by the map size. Let us also
recap that the above calculation breaks down at low wave numbers, and
thus for the too small values of $\theta$, since the GKS
effect would no longer be the dominant source of temperature
anisotropies. It is unclear at what $\ell$ this happens, but we note
that there is no sign of an $\ell^{-1}$ power law in \cite{Bevis:2007},
so the GKS effect may be a subdominant contribution to the 
cosmic string power spectrum for $\ell < 1500$.

\subsubsection{Collapsed triangles}
\label{sect:colconf}
There is however another flat triangle configuration for which all
wavenumbers scale similarly. For $\theta \to \pi$, we have a collapsed
triangle with $k_1=k_2=k$ and $k_3\to 2k$. Denoting by
\begin{equation}
\angcol \equiv \pi - \angsqz,
\end{equation}
the isosceles bispectrum reduces to
\begin{equation}
\BT_{k\angcol\angcol} \equiv \lim_{\angsqz \to \pi}
\BT_{\uiso}(k,\angsqz) \underset{\angcol\to 0}{\sim} +\epsilon^3 {3\pi}
 c_0\frac{\vb^2}{\tb^4}\frac{L\hat\xi}{\Area}
 \frac{1}{\hat\xi^2k^6} \dfrac{1}{\angcol^3}\,.
\end{equation}
Notice the divergence but with a positive bispectrum. Again, such a
divergence is not observable in a realistic situation due to the
finite detector resolution. Nevertheless, flat triangles should
provide the best framework for the search of a string bispectrum
signature. In fact, the change of sign in between the squeezed and
collapsed triangles is also of interest to improve the signal to noise
ratio. Substracting these two configurations with the appropriate
angles should enhance the string signal over the noise.

\subsubsection{Equilateral triangles}

For completeness, we give the bispectrum in the particular case
$\theta=\pi/3$, where the $\bk_a$ are arranged in an equilateral
triangle:
\begin{equation}
\left|\bk_1\right| =\left|\bk_2\right| = \left|\bk_3\right| =k .
\end{equation}
One has $\ka_{ab} = k^2/2$, and the factor in the squared brackets in
Eq.~(\ref{e:Beqn}) becomes $4/\sqrt{3}$. As the result,
\begin{equation}
\label{eq:beq}
 \BT_{kkk} \equiv \BT_{\ueq}(\bk_1,\bk_2,\bk_3) = -\epsilon^3 \frac{4\pi}{\sqrt{3}}
 c_0\frac{\vb^2}{\tb^4}\frac{L\hat\xi}{\Area} \frac{1}{\hat\xi^2k^6}\,.
\end{equation}
With $c_0>0$, such a configuration produces a negative bispectrum
decaying as $1/k^6$ but with an overall amplitude significantly
smaller than the collapsed and squeezed triangle configurations.

\section{Numerical estimate}
\label{sect:numerical}

In this section, we directly compute the three point function in
Fourier space from a set of $300$ CMB temperature maps induced by
cosmic strings in the flat sky approximation. These maps have been
obtained along the lines of Ref.~\cite{Fraisse:2007nu} from
Nambu--Goto numerical simulations in FLRW space-time. As detailed in
this reference, they are obtained from the GKS effect using
Eq.~(\ref{eTemGau}), and are therefore valid on small angular scales
only. In the next sections, we briefly recall the numerical method
used to generate the maps and then discuss our bispectrum estimator.
\subsection{Nambu--Goto numerical simulations}
Our FLRW numerical simulations are based on an improved version of the
Bennett and Bouchet Nambu-Goto cosmic string code~\cite{Bennett:1990,
 Ringeval:2007, Fraisse:2007nu}. The runs are performed in a comoving
box with periodic boundary conditions and whose volume has been scaled
to unity. The horizon size $\horizonini$ is a free parameter
controlling the initial string energy within a horizon volume. For the
simulations we performed, $\horizonini \simeq 0.185$. The string
network is assumed to come from Vachaspati-Vilenkin initial conditions
for which the long strings path is a random walk of correlation length
$\corrini$, plus a random transverse velocity component of root mean
squared amplitude $0.1$~\cite{Vachaspati:1984}. These parameters are
set as in Ref.~\cite{Fraisse:2007nu} to minimize the relaxation time
of the Vachaspati-Vilenkin string network toward its stable
cosmological configuration. The temperature maps are then produced
according the GKS effect generated by the strings
intercepting our past light cone, using the Eq.~(\ref{eTemGau})
\begin{equation}
\label{eq:stgsa}
\deTovT \simeq 
 \dfrac{8\pi\mathrm{i}\, \newton \tension}{\vect{\kperp}^2}
 \int_{\vect{X}\,\cap\,\vect{x}_\gamma}
 \left(\vect{u} \cdot \vect{\kperp}\right)
 e^{-\mathrm{i}\, \vect{\kperp} \cdot {\vect{X}}}\,
 e^{-\tau}\epsilon\, \ud \sigma\,,
\end{equation}
where $\tau(\bX)$ is the optical depth to the position of the string,
along the line of sight $\unitn$, and
\begin{equation}
\label{eq:udef}
\vect{u} = \dot{\vect{X}} - 
 \dfrac{(\unitn \cdot \vect{X}') \cdot \vect{X}'}{1 +
 \unitn \cdot \dot{\vect{X}}}.
\end{equation}
The cosmic string simulations are used to compute $\vect{u}$, based on
the string trajectories $\vect{X}$. The map generation procedure
introduces one additional parameter which is the redshift at which we
start the simulations $z_\ui$. It has been set to the last scattering
surface, namely $z_\ui=1089$, and in a flat ``Lambda-Cold-Dark-Matter''
($\LCDM$) universe, using fiducial values for the density parameters
compatible with the three-year Wilkinson Microwave Anisotropy Probe
(WMAP) data~\cite{Spergel:2007}, this corresponds to a numerical
comoving box of $\boxmpc\simeq 1.7\,\Gpc$. Such a size subtends a
angle of $\angfov \simeq 7.2^\circ$ in the sky.  The Nambu-Goto
simulations and the associated CMB maps therefore depend on only two
parameters: the string energy per unit length $\tension$, and the
initial correlation length $\corrini$. The dependence on $\corrini$
should drop out at late times, as the network is believed to approach
a self-similar or scaling configuration.  However, in a real
Nambu-Goto simulation, there is still some non-scaling structure
evident on the smallest length scales~\cite{Fraisse:2007nu}.
\begin{figure}
\begin{center}
\includegraphics[width=8.cm]{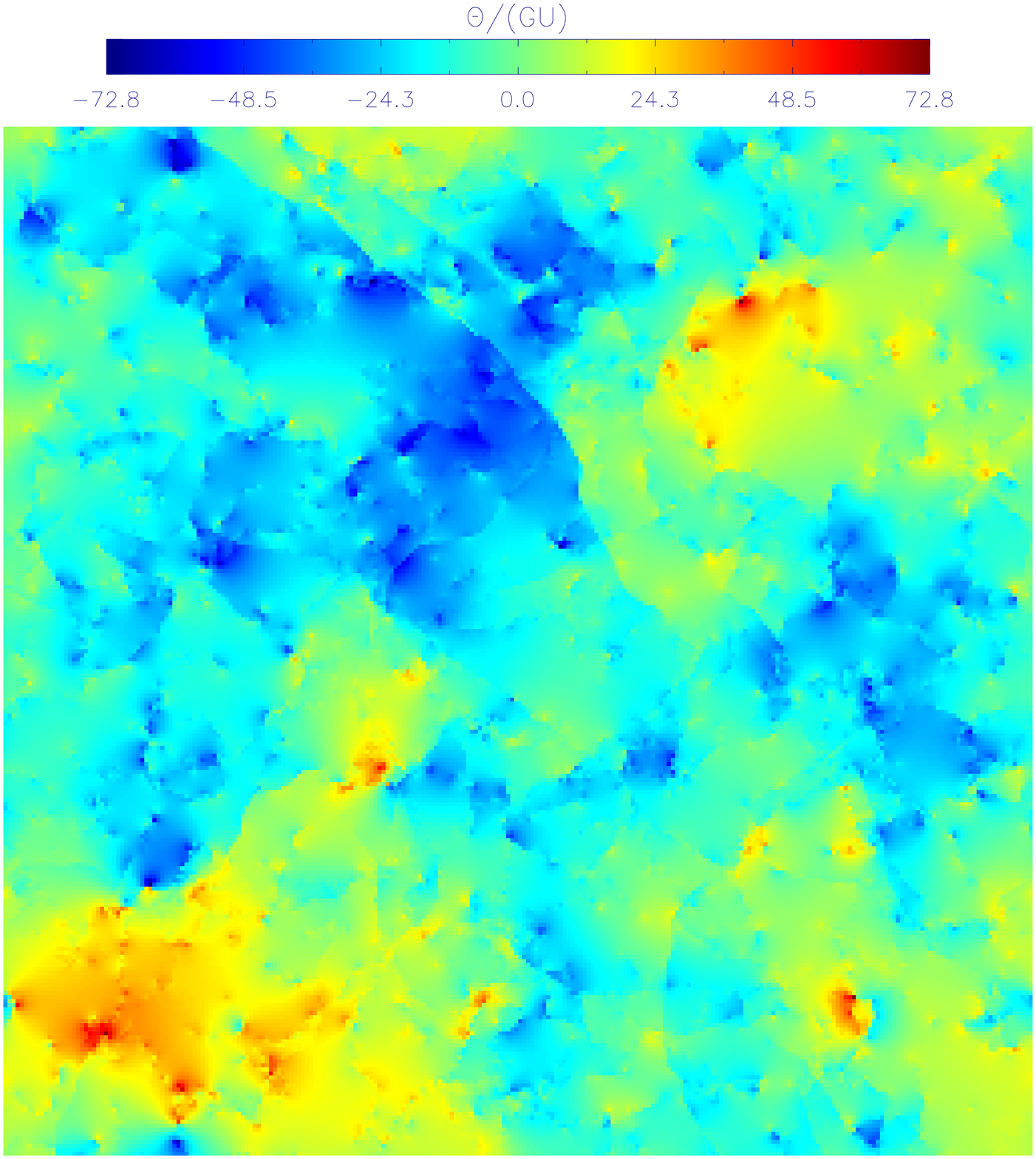}
\includegraphics[width=8.cm]{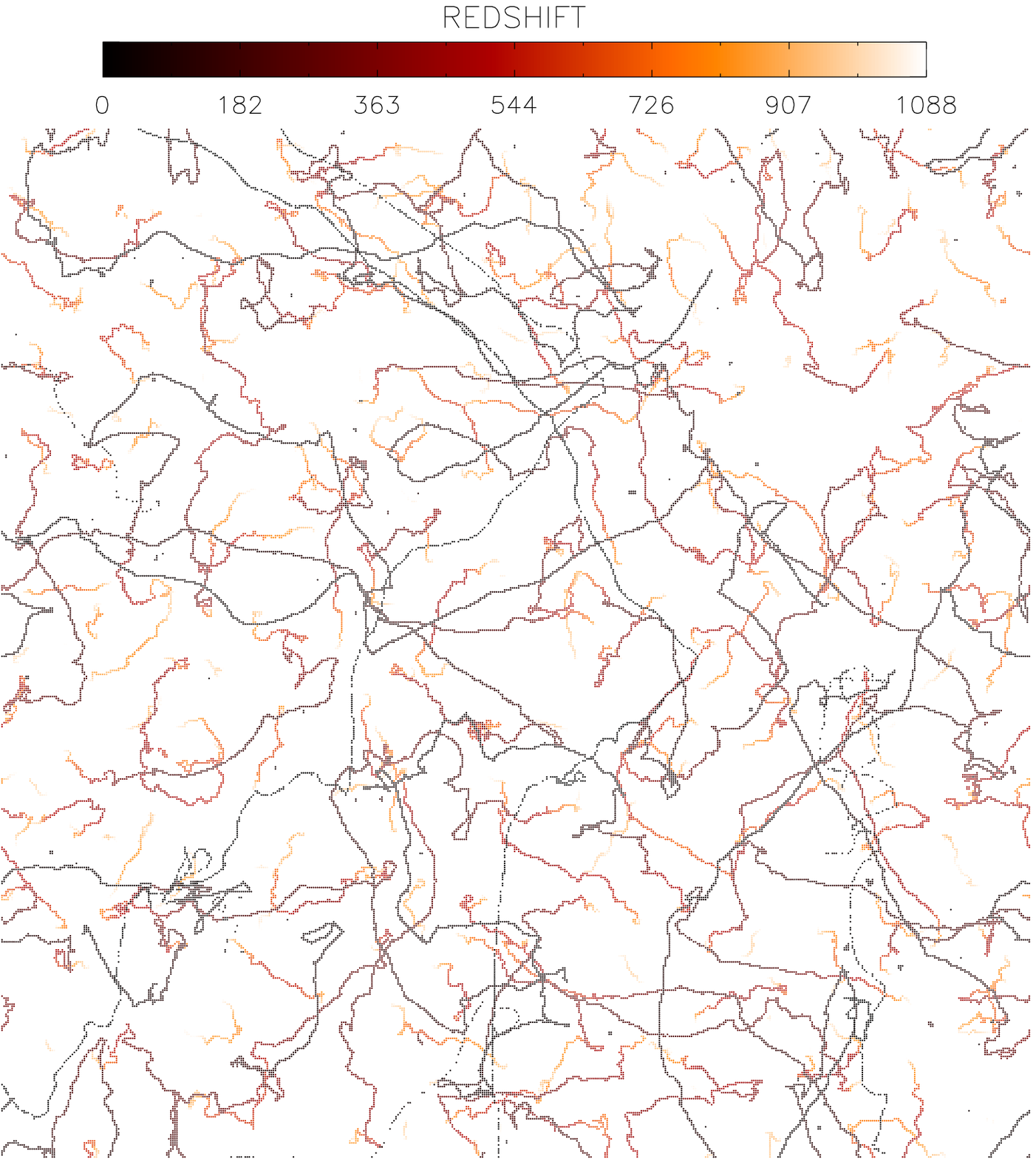}
\caption{Typical CMB temperature map on a $7.2^\circ$ field
 (resolution of $\angres=0.42'$, $\npixel^2=1024^2$) obtained from
 Nambu--Goto cosmic string simulations (upper panel). The lower panel
 traces back the strings projected on our past light cone.}
\label{fig:smap}
\end{center}
\end{figure}
As an illustration example, a typical CMB temperature anisotropy map
is represented in Fig.~\ref{fig:smap}, together with the seeding
strings projected on our past light cone. We have used our numerical
simulations to create $300$ statistically independent temperature maps
from which one can construct a bispectrum estimator.

\subsection{Reduced bispectrum estimator}

\subsubsection{Scale convolution method}
\begin{figure}
\begin{center}
\includegraphics[width=8.cm]{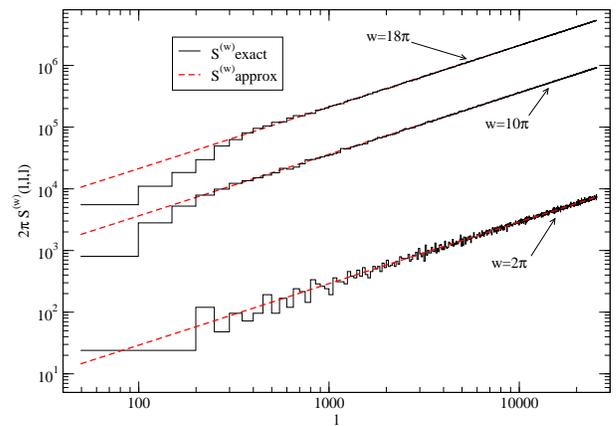}
\caption{Exact versus approximated geometrical $S^{(w)}_{\ell \ell
   \ell}$ factors for a map with $1024^2$ pixels and various choices
 of the window function with $w$ ($w=2\pi$ corresponds to one
 Fourier mode per bandwidth). For large values of $w$, the deviations
 for the small multipoles come from the averaging effect. For small
 angles we are interested in, Eq.~(\ref{eq:appwig}) provides a good
 approximation to the exact expression (\ref{eq:invwig}).}
\label{fig:wigtest}
\end{center}
\end{figure}

The bispectrum computations use the scale convolution method
introduced in Ref.~\cite{Spergel:1999xn} and applied to the flat sky
approximation in Ref.~\cite{Aghanim:2003fs}. This method relies on the
choice of unity window functions in Fourier space $\window{u}{\kperp}$
peaked around a particular wavenumber $u$.  Defining
\begin{equation}
 \deTovT_{u}(\vect{x}) \equiv \int\dfrac{\ud \vect{\kperp}}{(2\pi)^2} \FFTheta{\vect{\kperp}}
 \window{u}{\kperp} e^{-i \vect{\kperp} \cdot \vect{x}},
\end{equation}
one can construct an estimator of the three point function in Fourier
space by remarking that
\begin{equation}
\begin{aligned}
 \int & \deTovT_{k_1}(\vect{x}) \deTovT_{k_2}(\vect{x})
 \deTovT_{k_3}(\vect{x}) \ud \vect{x} = \int \dfrac{\ud \vect{p}\ud
   \vect{q} \ud \vect{k}}{(2\pi)^6}
 \FFTheta{\vect{p}}\FFTheta{\vect{q}} \FFTheta{\vect{k}} \\ & \times
 \window{k_1}{p}\window{k_2}{q}\window{k_3}{k} (2\pi)^2
 \delta(\vect{p} + \vect{q} + \vect{k}).
\end{aligned}
\end{equation}
For peaked enough window functions, $\FFTheta{\vect{k}}$ remains
constant over the window functions width and we construct our reduced
bispectrum estimator as
\begin{align}
\label{eq:rbisDef}
\rbis_{k_1k_2k_3} = \frac{1}{S^{(w)}_{k_1 k_2 k_3}}
\lvev{\int \deTovT_{k_1}(\vect{x})
 \deTovT_{k_2}(\vect{x}) \deTovT_{k_3}(\vect{x}) \ud \vect{x}}.
\end{align}
where the function $S^{(w)}$ is the flat sky equivalent of the inverse Wigner-3j
symbols and reads
\begin{align}
\label{eq:invwig}
S^{(w)}_{k_1 k_2 k_3} = \int \dfrac{\ud \vect{p} \ud
 \vect{q}}{(2\pi)^4} \window{k_1}{p} \window{k_2}{q}
\window{k_3}{\left|\vect{p}+\vect{q}\right|}.
\end{align}
As noted in Ref.~\cite{Aghanim:2003fs}, $S$ is of geometrical nature
and needs to be computed only once. However, at small angular scales,
$S$ is generically a four-dimensional integral whose computation can
be time consuming for the large wavenumbers. For thin enough window
functions, it is nevertheless possible to derive an analytical
approximation. Assuming that $\window{u}{k}=1$ for $u-w/2 < k <
u+w/2$, and zero otherwise, for small enough width $w$ one has
\begin{equation}
\window{u}{k} \simeq w \delta(k-u).
\end{equation}
With $\vect{k_1}$, $\vect{k_2}$ and $\vect{k_3}$ forming a triangle,
Eq.~(\ref{eq:invwig}) can be exactly integrated and one finds
\begin{equation}
\label{eq:appwig}
S^{(w)} \simeq \left( \dfrac{w}{2 \pi}\right)^3 \dfrac{4 k_1 k_2
 k_3}{\sqrt{\left[ \left(k_1 + k_2\right)^2 - k_3^2\right]
   \left[k_3^2 - \left(k_1-k_2\right)^2\right]}}\,.
\end{equation}
As can be checked in Fig.~\ref{fig:wigtest}, the analytical
approximation of $S^{(w)}_{k_1 k_2 k_3}$ given in
Eq.~(\ref{eq:appwig}) is particularly accurate for the large
wavenumbers.
This approach ends up being numerically convenient since it requires
only three Fourier transforms to compute the $\deTovT_{k}(\vect{x})$,
together with an integration over all map pixels.

\subsubsection{Statistical averaging and numerical tests}
From the previous discussion, the bispectrum is extracted and averaged
over the $300$ statistical independent maps and for the various
triangle configurations discussed in Sect.~\ref{sec:triconfig}. Our
results are presented in the next section. The variance of the
bispectrum over the different maps is used to provide an estimate of
the statistical errors associated with our approach. Notice however,
as discussed at length in Ref.~\cite{Fraisse:2007nu}, the Nambu-Goto
string simulation do also lead to some systematic errors due to the
non-scaling structure. This can be greatly reduced by eliminating
some of the small loops from the network.
\begin{figure}
\begin{center}
\includegraphics[width=8.cm]{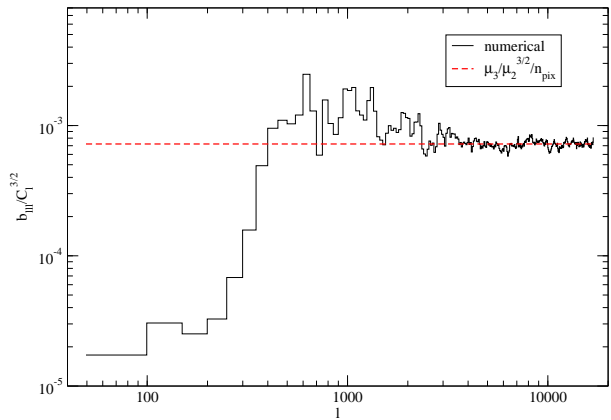}
\caption{The reduced bispectrum computed by the scale convolution
 method and averaged over a hundred of non-Gaussian synthetic $\LCDM$
 temperature maps whose probability distribution function is given by
 Eq.~(\ref{eq:ngpdf}). The red dashed line is the expected
 non-Gaussian signal. The missing power for low multipoles comes from
 the cutoff at  $\ell_w\simeq w/\angfov\simeq 1050$ introduced by the choice
 of large $w$ values for the window functions.}
\label{fig:bngcdm}
\end{center}
\end{figure}
In order to check our numerical bispectrum estimator, we have
performed two tests. The first was to generate a set of synthetic
non-Gaussian CMB temperature maps from the non-Gaussian probability
distribution
\begin{equation}
\label{eq:ngpdf}
P_{\alpha,\sigma}(\deTovT) = \dfrac{e^{-\deTovT^2/(2
   \sigma^2)}}{\sqrt{2\pi}
 \sigma}
\left[ \sqrt{1-\alpha^2}  + \dfrac{\alpha}{\sqrt{48}}
 H_3\left(\dfrac{\deTovT}{\sqrt{2}\sigma}\right) \right],
\end{equation}
where $H_3$ stands for the third Hermite polynomial. As shown in
Ref.~\cite{Contaldi:2001wr, Rocha:2005}, such a statistics leads to a
reduced bispectrum whose amplitude is given
\begin{equation}
\label{eq:bhermite}
\dfrac{b_{\ell_1\ell_2\ell_3}}{\sqrt{C_{\ell_1} C_{\ell_2} C_{\ell_3}}}
 = \dfrac{1}{\npixel} \dfrac{\mu_3}{\mu_2^{3/2}}\,,
\end{equation}
where $\mu_2=\sigma^2(1+6\alpha^2)$ and $\mu_3=
(2\sigma^2)^{3/2}\alpha \sqrt{3(1-\alpha^2)}$ are the second and third
central moment, respectively. It is not difficult to show that
$\mu_3/\mu_2^{3/2}$ is maximised for $\alpha^2=(7-\sqrt{43})/6$. In
Fig.~\ref{fig:bngcdm}, we have plotted the bispectrum obtained from
the scale convolution method averaged over a hundred of such
non-Gaussian $\LCDM$ temperature maps together with its analytical
expectation. At small scales, and up to the statistical errors, the
non-Gaussian signal is recovered with the right amplitude. Notice that
the loss of power for the low multipoles is simply an artifact coming
from the choice of a large window function width ($w=42\pi$). This
necessarily introduces a lower frequency cutoff around 
$\ell_w\simeq w/\angfov\simeq 1050$, and also close to the Nyquist
frequency (not visible on the plots). As discussed in the following,
increasing the values of $w$ reduces the variance of the bispectrum
estimator\footnote{Simply by increasing the number of equivalent
 triangle configurations in Fourier space probed by the window
 functions.} at the price of losing information for the lower
multipoles and those close to the Nyquist frequency.  The second test
we have performed is to integrate the cosmic string bispectrum over
all possible triangle configurations such as to check that we indeed
recover the skewness of the string temperature anisotropies. Averaged
over the $300$ string maps, we find the mean sample skewness to be
negative
\begin{equation}
\label{eq:skew}
g_1 \equiv \left \langle
\frac{\overline{(\deTovT-\bar{\deTovT})^3}}{\sigma^3} \right \rangle
\simeq -0.22 \pm 0.12,
\end{equation}
where the brackets stand for the mean over different realisations
while the bar denotes averaging on each map. The variance itself
averages to
\begin{equation}
\label{eq:var}
\sigma^2 = \left\langle \overline{(\deTovT -
 \bar{\deTovT})^2}\right \rangle\simeq   \left(150.7 \pm 18 \right) (\GU)^2.
\end{equation}
The quoted errors are statistical and refer to the square root of the
variance between the different realisations.

\subsection{Flat triangle configurations}

The previously described method has been applied to extract the
isosceles bispectra in the flat triangle configurations for which the
amplitude is maximal.

\subsubsection{Squeezed triangle}

\begin{figure}
\begin{center}
\includegraphics[width=8.cm]{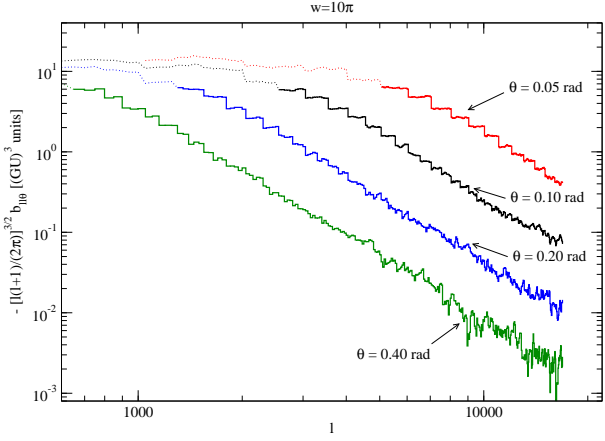}
\caption{Squeezed angle dependency of the mean squeezed bispectrum
 $\left[\ell(\ell+1)/(2\pi)\right]^{3/2} b_{\ell\ell\angsqz}$. The
 spurious plateau (dotted lines) for the lower multipoles comes from
 the cutoff associated with a given squeezed angle $\angsqz$ together
 with the window function width $w$ and occurs at
 $\ell_{\min}=w/(\angfov \angsqz)$.}
\label{fig:bllthw5}
\end{center}
\end{figure}

Let us first consider the triangle configuration in the squeezed limit
such that
\begin{equation}
\label{eq:lltconf}
\kperp_1 =\kperp_2=\kperp, \quad \kperp_3 \simeq \kperp \theta,
\end{equation}
that we refer to as ``$\ell\ell\theta$''. The mean value of
$b_{\ell\ell \theta}$ over the different maps have been plotted in
Fig.~\ref{fig:bllthw5} and for various values of the squeezed angle
$\angsqz$. As found analytically, the bispectrum is negative while the
overall amplitude is found to be enhanced by a factor
$1/\angsqz^3$.  We have used a window function width equals to
$w=10\pi$ for this plot. However, due to the finite field of view,
we cannot go to very small squeezed angle without truncating the lower
modes. From Eq.~(\ref{eq:lltconf}), the lowest multipole achievable is
therefore given by 
\begin{equation}
\label{eq:lmin}
\ell_{\min}= \dfrac{w}{\angfov\angsqz}\, ,
\end{equation}
where the window function cutoff has also been included.

\begin{figure}
\begin{center}
\includegraphics[width=8.cm]{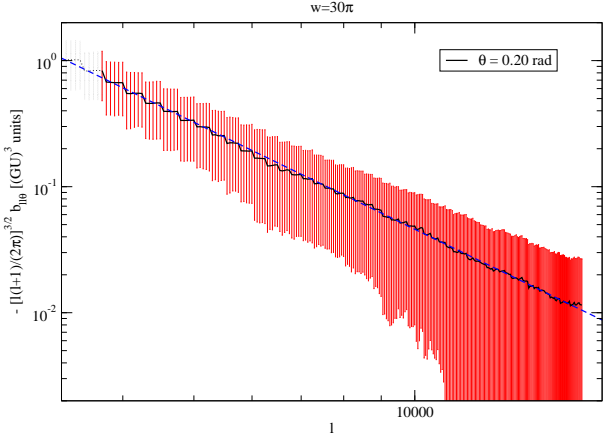}
\caption{Mean value and standard deviation of the squeezed bispectrum
 $\left[\ell(\ell+1)/(2\pi)\right]^{3/2} b_{\ell\ell\angsqz}$ for
 $\angsqz=0.2$ radians. The window function width has been increased
 to $w=30\pi$ to reduce the variance pushing up the lower multipole
 cutoff to $\ell_{\min}=w/(\angfov\angsqz)\simeq 3750$. The dashed
 line is the best power law fit whose power has been set to the one
 obtained from the equilateral configuration [see
   Eq.~(\ref{eq:blllfit})].}
\label{fig:bllthw15}
\end{center}
\end{figure}

In order to determine the power law behaviour of the bispectrum , we
have plotted in Fig.~\ref{fig:bllthw15}, the mean value and standard
deviation over the $300$ different maps of the squeezed bispectrum for
the particular angle $\angsqz=0.2$ radians. The variance has been
reduced as far as we could by using a large window function width
$w=30\pi$. This explains the low multipole cutoff at a quite large
value of $\ell_{\min}\simeq 3750$.  A power law fit against the
mean numerical estimator for $\ell>\ell_{\min}$ and truncated at
$\ell_{\max}=16000$ gives
\begin{equation}
\label{eq:blltfit}
\left[\ell(\ell+1)\right]^{3/2} b_{\ell\ell\angsqz} \underset{\ell \gg
 1}{\propto} \ell^{-q} \quad \mathrm{with} \quad q = 2.82\,.
\end{equation}
We do not quote error bars since this value is relatively sensitive to
the choice of $\ell_{\max}$ and, as can been seen in
Fig.~\ref{fig:bllthw15}, the standard deviation of our estimator
becomes an order of magnitude bigger than the bispectrum itself for
$\ell>16000$.

The power dependency of Eq.~(\ref{eq:blltfit}) is remarkably close to
the analytical expectation. In fact, the non-integer value for the
power may be expected due to the fractal properties of the Nambu--Goto
strings~\cite{Martins:2006}. As shown in Ref.~\cite{Fraisse:2007nu},
the power spectrum itself was found to behaves as
\begin{equation}
\label{eq:clsfit}
\ell(\ell+1) C_\ell \underset{\ell \gg 1}{\propto} \ell^{-p} \quad
\mathrm{with} \quad p = 0.889\,.
\end{equation}
From Eqs.~(\ref{eq:blltfit}) and (\ref{eq:clsfit}), we the find the
analytical power law behaviour of the ratio
\begin{equation}
\label{eq:bllloclfit}
\dfrac{b_{\ell\ell\angsqz}}{C_\ell^{3/2}} \underset{\ell \gg 1}{\propto}
 \ell^{-(q-3p/2)} \quad \mathrm{with} \quad q-\dfrac{3}{2}p = 1.49\,.
\end{equation}
The dashed blue line in Fig.~\ref{fig:bllthw15} is the best power law
fit $\ell^{-q}$.

\begin{figure}
\begin{center}
\includegraphics[width=8.cm]{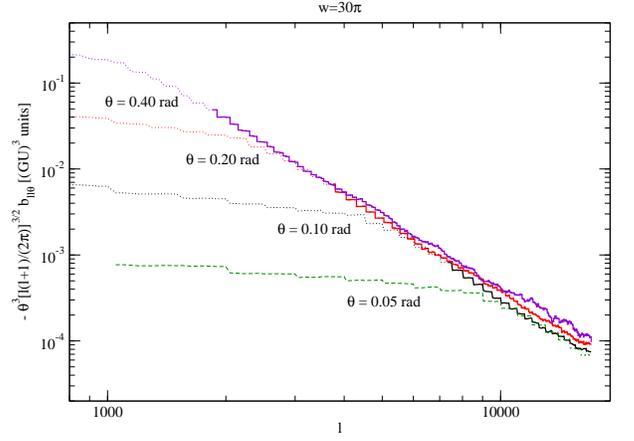}
\caption{Rescaled mean squeezed bispectrum $\angsqz^3
 [\ell(\ell+1)/(2\pi)]^{3/2} b_{\ell\ell\angsqz}$ showing the
 $1/\angsqz^3$ dependency. Notice the large window function width
 $w=30\pi$ chosen to reduce the statistical errors and pushing up the
 numerical multipole cutoff $\l_{\min}$.}
\label{fig:bllttheta3}
\end{center}
\end{figure}

The $1/\angsqz^3$ dependency of the squeezed bispectrum is illustrated
in Fig.~\ref{fig:bllttheta3}. Again, to reduce the variance, the width
has been set to $w=30\pi$ and we have plotted $\angsqz^3
[\ell(\ell+1)/(2\pi)]^{3/2} b_{\ell\ell\angsqz}$ for the four
$\angsqz$ values of Fig.~\ref{fig:bllthw5}. As expected, all rescaled
mean values match above their respective multipole cutoff
$\ell_{\min}$  [see Eq.~(\ref{eq:lmin})].

Concerning the overall amplitude, it can be evaluated around the
minimum variance multipole and we find at $\ell=5000$
\begin{equation}
\label{eq:blltamp}
\left[\ell(\ell+1)/(2\pi)\right]^{3/2} b_{\ell\ell\angsqz} \simeq \left(-2.7
\pm 1.4\right) \times 10^{-3} \left(\dfrac{\GU}{\angsqz}\right)^3.
\end{equation}

To conclude this section, the squeezed bispectrum is found to follow
the analytical results of Sect.~\ref{sect:analytical}, up to the
non-integer power law behaviour with respect to the multipole moment
which may be interpreted as a fractal effect associated with the
Nambu--Goto string small scale structure.

\subsubsection{Collapsed triangles}

As discussed in Sect.~\ref{sect:colconf}, the collapsed triangles
correspond to the limit $\angcol\to 0$ for the isosceles configuration
and to the mode values
\begin{equation}
k_1=k_2=k, \qquad k_3 \simeq 2k.
\end{equation}
The scale convolution method is hardly applicable in this case since
all collapsed triangles fitting in the window function width will be
indistinguishable. Although this averaging effect allowed us to
increase the statistics for the squeezed triangle configurations, for
the collapsed ones one cannot longer assume that the bispectrum is
constant over the window function, especially since it diverges in the
limit $\angcol\to 0$. As a result, $w$ has been kept to a low value
($w=3\pi$) and the modes have been binned to reduce the statistical
errors.
\begin{figure}
\begin{center}
\includegraphics[width=8.5cm]{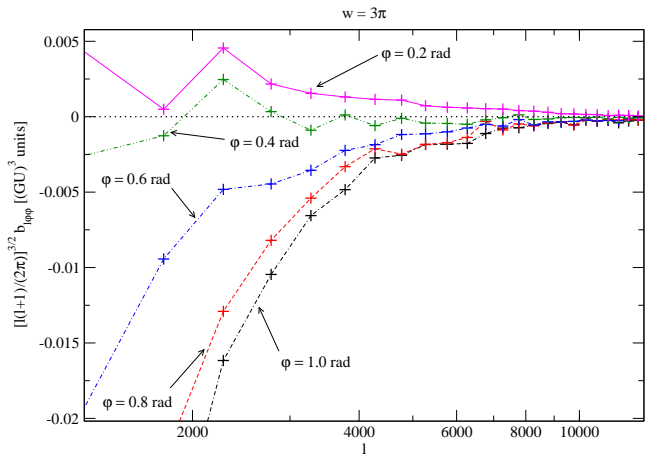}
\caption{Mean collapsed bispectrum $ [\ell(\ell+1)/(2\pi)]^{3/2}
 b_{\ell\angcol\angcol}$. This configuration is extremely sensitive
 to the window function properties and statistical errors
 dominate. The change of sign for small values of $\angcol$ is
 nevertheless observable.}
\label{fig:blpp}
\end{center}
\end{figure}
As can be seen in Fig.~\ref{fig:blpp}, our numerical
estimation does not allow to accurately traces the bispectrum down to
the small values of $\angcol$. Nevertheless, we do observe the change
of sign for $\angcol \to 0$ albeit at a smaller values than the one
predicted analytically.

\subsection{Equilateral configurations}

\begin{figure}[h]
\begin{center}
\includegraphics[width=8.cm]{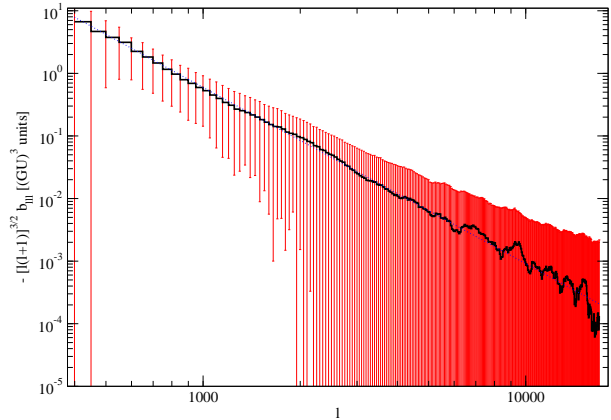}
\caption{Mean and standard deviation of the equilateral bispectrum
 $\left[\ell(\ell+1)/(2\pi)\right]^{3/2}b_{\ell\ell\ell}$ over our
 $300$ string temperature maps and obtained for window function
 width $w=30\pi$. The blue dotted line is the power law fit of
 Eq.~(\ref{eq:blllfit}).}
\label{fig:blllhw15}
\end{center}
\end{figure}

The previously described method has also been applied to extract the
equilateral bispectrum $b_{\ell\ell\ell}$. In Fig.~\ref{fig:blllhw15},
we have plotted its mean value over the $300$ different maps as well
as its standard deviation. To reduce the variance of the estimator we
have chosen $w=30\pi$ which gives a low multipole cutoff $\ell_w\simeq
750$. The equilateral bispectrum being of smaller amplitude that the
squeezed one, its numerical determination is quite challenging. It is
however clear from this plot that $b_{\ell\ell\ell} < 0$ in agreement
with the analytical results of Sect.~\ref{sect:analytical}.

\begin{figure}
\begin{center}
\includegraphics[width=8.cm]{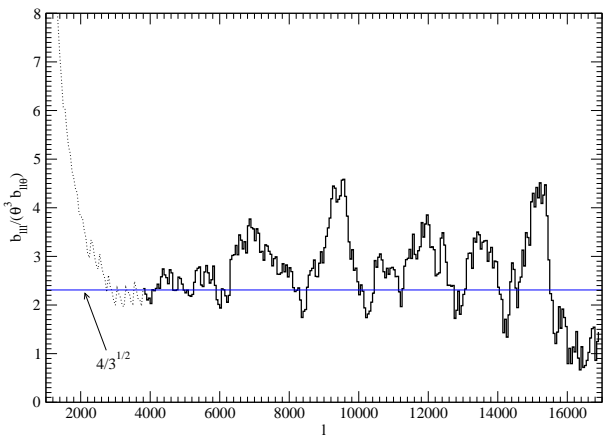}
\caption{Ratio of the mean bispectra $b_{\ell\ell\ell}/(\angsqz^3
   b_{\ell\ell\angsqz})$ obtained for $\angsqz=0.2$ radians. Up to
 the statistical errors, the expected value $4/\sqrt{3}$ is recovered
 for $\ell>\ell_{\min}$.}
\label{fig:blllobllt}
\end{center}
\end{figure}

A power law fit against the mean numerical estimator over the range of
multipoles $\ell_w <\ell < \ell_{\max}$, with $\ell_{\max}=16000$
gives the same power as for the squeezed triangle configurations,
namely
\begin{equation}
\label{eq:blllfit}
\left[\ell(\ell+1)\right]^{3/2} b_{\ell\ell\ell} \underset{\ell \gg
 1}{\propto} \ell^{-q} \quad \mathrm{with} \quad q = 2.82\,.
\end{equation}
Concerning the overall amplitude, we find at $\ell=1000$
\begin{equation}
\label{eq:blllamp}
\left[\ell(\ell+1)/(2\pi)\right]^{3/2} b_{\ell\ell\ell} \simeq
\left(-0.53 \pm 0.3\right) (\GU)^3,
\end{equation}
which is again of the same order of magnitude as our analytical
estimate. Using the power law fit, such an amplitude can be
extrapolated up to $\ell=5000$ and compared to
Eq.~(\ref{eq:blltamp}). In particular, we recover, up to the
statistical errors, the analytical result
\begin{equation}
\dfrac{b_{\ell\ell\ell}}{\angsqz^3 b_{\ell\ell\angsqz}} = \dfrac{4}{\sqrt{3}}\,.
\end{equation}
For completeness, the above ratio has been plotted for $\angsqz=0.2$
radians in Fig.~\ref{fig:blllobllt}.

 As can be seen in Fig.~\ref{fig:blllhw15}, but also in
Fig.~\ref{fig:bllthw15}, the bispectrum decays faster than its
standard deviation at small scales. This can be understood by using
the nearly Gaussian approximation in which the non-Gaussian
trispectrum is neglected. Under these assumptions, the variance of
the bispectrum estimator, for the equilateral configurations, is given
by~\cite{Aghanim:2003fs, Bartolo:2004if}
\begin{equation}
\sigma_b^2 = \dfrac{6}{S_{\ell\ell\ell}^{(w)}} C_\ell^{3}.
\end{equation}
Using Eq.~(\ref{eq:appwig}), one gets
\begin{equation}
\label{eq:sigblll}
\sigma_b =\sqrt{\dfrac{2\pi}{\angfov}}
\left(\dfrac{2\pi}{w}\right)^{3/2} \left[\dfrac{27}{4
   \ell(\ell+1)}\right]^{1/4} \left[\dfrac{\ell(\ell+1) C_\ell}{2 \pi
 }\right]^{3/2}.
\end{equation}
As expected, the variance is reduced by a factor $w^3$ coming from the
different numbers of equivalent triangle configurations probed by
window functions of width $w$. From Eqs.~(\ref{eq:clsfit}) and
(\ref{eq:sigblll}), one gets $\sigma_b\propto \ell^{-4.83}$ which
vanishes slower than the bispectrum estimator at large multipoles.
\begin{figure}
\begin{center}
\includegraphics[width=8.cm]{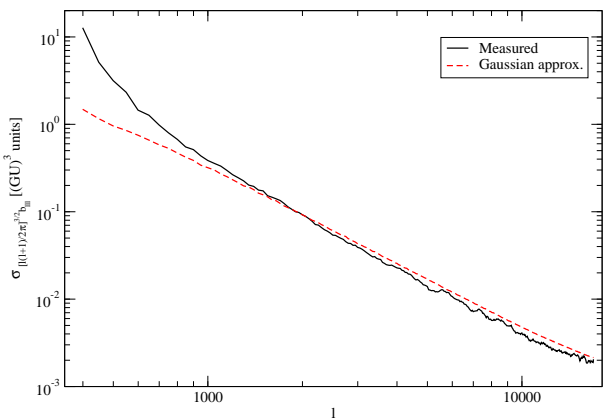}
\caption{Standard deviation of the bispectrum estimator over the $300$
 string temperature maps (black solid curve) compared with the nearly
 Gaussian random expected standard deviation obtained from
 Eq.~(\ref{eq:sigblll}).}
\label{fig:sigblll}
\end{center}
\end{figure}
In Fig.~\ref{fig:sigblll}, we have plotted the standard deviation
obtained from Eq.~(\ref{eq:sigblll}), and relying only on the mean
string power spectrum, together with the measured standard deviation
of our bispectrum estimator. There is good agreement for both the
power law behaviour and the overall amplitude.  From this section, we
conclude that although the cosmic string equilateral bispectrum is
non-vanishing, its measurement at small angles may be difficult due to
its fast decay compared to the associated standard deviation. This is
also the case for the squeezed and collapsed configurations, but the
formers being of higher amplitude, this should be less problematic.

\section{Comparison with data}

From both the analytical and numerical results, the cosmic string
bispectra associated with the flat triangle configurations are the
best suited to look for string signatures. However, it is not easy to
compare with existing bispectrum estimates as much of the literature
focuses on particular local models of primordial non-Gaussianity in
which the Newtonian potential is parametrised~\cite{Komatsu:2001rj}
\begin{equation}
 \Phi(\bx) = \Phi(\bx)_\uL + \fNLloc\left[\Phi^2_\uL(\bx) - \lvev{\Phi^2_\uL(\bx)}\right].
\end{equation}
The local type of primordial non-Gaussianity produces a bispectrum
whose maximal amplitude occurs for squeezed triangle configurations,
as it is the case for the cosmic strings. However, as a result of the
CMB transfer functions, a given value of $\fNLloc$ corresponds to
oscillating damped patterns of the CMB temperature bispectrum, which
are therefore completely different from the power law we have derived
for the strings at small scales. The current bounds on $\fNLloc$ being
precisely obtained from template matching procedures, they cannot be
applied to the string bispectrum~\cite{Komatsu:2008hk,
 Smith:2009jr}. Notice that the analysis concerned with the other
type of primordial non-Gaussianities, such as the equilateral ones
characterised by $\fNLeq$, are also affected by the CMB transfer
function~\cite{Smith:2009jr} and cannot straightforwardly be compared
to our findings.

However, a simple calculation we can do at this stage is, at a given
angular scale, to compare the overall amplitude the CMB bispectrum
induced by a primordial $\fNLloc$ with the one induced by the presence
of cosmic strings. For simplicity, let us choose the equilateral
configuration (which is not of maximal amplitude for both the strings
and local primordial non-Gaussianity). With COBE-normalised CMB
temperature fluctuations, and neglecting Silk damping,
Ref.~\cite{Komatsu:2001rj} gives the amplitude of the oscillating CMB
bispectrum induced by primordial non-Gaussianity:
\begin{equation}
\label{eq:btlllprim}
\max{|\BT_{lll}|} \simeq (2\times 10^{-17})l^{-4}  |\fNLloc|.
\end{equation}
We can now estimate the size of the equilateral bispectrum from
strings if they contribute about $10\%$ of the temperature power
spectrum at $\ell=10$. In this case $\epsilon \simeq 2\times 10^{-5}$,
and assuming Abelian Higgs strings where $G\tension \simeq
0.7\times10^{-6}$~\cite{Bevis:2007gh,Bevis:2006mj}, one has
\begin{equation}
 \BT_{kkk} \simeq - (4\times 10^{-14})c_0
 \frac{\vb^2}{\tb^4}\frac{L\hat\xi}{\Area}
 \frac{1}{\hat\xi^2k^6} \(\frac{G\tension}{0.7\times 10^{-6}}\)^3.
\end{equation}
Matching the two suggests that string fluctuations may generate a
temperature bispectrum at $\ell \simeq 500$ as large as the one that
would be produced by a local $|\fNLloc|$ of around $10^3$ ($k\hat\xi
\simeq 1$ is near the peak of the scalar modes at $\ell \sim
500$). This number should be considered as an order of magnitude
estimate since the string bispectrum calculation is only accurate for
higher $\ell$, while Eq.~(\ref{eq:btlllprim}) is applicable for $\ell
\le 1000$~\cite{Komatsu:2001rj}. Let us also mention that this value
for $\fNLloc$ is only valid at $\ell\simeq 500$ and does not represent
the value that would be obtained by a local primordial bispectrum
template matching against the string bispectrum. In order to obtain
such a value, one should perform a Fisher matrix analysis over all the
multipoles along the lines of Refs.~\cite{Sefusatti:2007ih,
 Nitta:2009jp}. Such an analysis is outside the scope of this paper
and we leave it for a forthcoming work.  
\section{Conclusion}

We have analytically and numerically derived the CMB temperature
bispectrum induced by cosmic strings from the Gott--Kaiser--Stebbins
effect. This effect is the main source of string induced CMB
anisotropies at small angle and our result is applicable at those
angular scales only.  The bispectrum is negative for equilateral and
all isosceles squeezed configurations, having a power law decay close
to $\ell^{-6}$.  The squeezed isosceles bispectrum is significantly
amplified for small squeezing angle by a factor
$\angsqz^{-3}$. Although the vanishing limit is forbidden by any
finite detector resolution, one should expect the ongoing and future
high resolution CMB experiment to be more sensitive to these
signatures~\cite{Ruhl:2004,Huffenberger:2005,Ami:2006}. The collapsed
isosceles configuration still exhibits the same power law, an angle
dependency in $\angcol^{-3}$ (with $\angcol=\pi-\angsqz$) but leads to
a positive bispectrum. This feature could allow an enhancement of the
string signal by subtracting the bispectrum of these two
configurations.   The numerically measured amplitudes have been
given in the text and, at the particular scale $\ell=500$, they
roughly correspond to a $|\fNLloc| \simeq 10^3$ for the local type of
primordial non-Gaussianity in the equilateral configurations. This
number is of illustrative purpose only and should be treated with
caution. Firstly, the temperature bispectrum induced by the strings
has a different multipole dependency than the one coming from
primordial non-Gaussianities. As a result, the value for $\fNLloc$
quoted above is definitely not the one that would be obtain by
applying a primordial bispectrum template matching over the actual
string bispectrum. In particular, the current bounds derived so far on
the primordial $\fnl$ values do not apply for the strings. Moreover,
even at the string prediction level, there are competing and
correlated effects at this angular scale, notably the acoustic
oscillations that we have not considered here. Nevertheless, this
value for $\fNLloc$ shows that this is an interesting angular scale to
search for string non-Gaussianity and that devoted string template
matching should be used in the search of CMB non-Gaussian signals.

Finally, as the power law dependence of the bispectrum variance may
suggest, we should expect an even stronger non-Gaussian signal from
the cosmic string trispectrum and we leave its analysis for a
forthcoming work.

\acknowledgments This work is partially supported by the Belgian
Federal Office for Scientific, Technical, and Cultural Affairs through
the Inter-University Attraction Pole Grant No. P6/11. The cosmic
string simulations have been performed thanks to computing support
provided by the Planck-HFI processing center at the Institut
d'Astrophysique de Paris.

\appendix

\section{String coordinate integrations}
\label{sec:strginteg}
In this Appendix we perform the integrals over the string coordinates
$\si_a$ in the expression (\ref{e:Beqn}) for the bispectrum. We repeat
the expressions for convenience:
\begin{equation}
\begin{aligned}
 \BT(\bk_1,\bk_2,\bk_3) & = -
 \epsilon^3\frac{1}{\Area}\frac{k_{1_A}k_{2_B}k_{3_C}}{k_1^2k_2^2k_3^2} \int
 \ud\si_1 \ud\si_2 \ud\si_3 \\ & \times \lvev{C^{ABC}D} e^{-\half \vev{D^2}}.
\end{aligned}
\end{equation}
where 
\begin{equation}
\begin{aligned}
& k_{1_A}k_{2_B} k_{3_C}
\lvev{C^{ABC} D} e^{-\half \vev{D^2}}  = \\ \frac14 \bigg\{ 
&\ka_{12}\[\ka_{13}\Pi( \si_{13}) + \ka_{23} \Pi( \si_{23}) \] V(\si_{12}) \\
+ &\ka_{13} \[  \ka_{12}\Pi( \si_{12})  + \ka_{23} \Pi(\si_{32}) \] V(\si_{31})  \\
+ &\ka_{23}\[  \ka_{12}\Pi( \si_{21}) +  \ka_{13} \Pi( \si_{31}) \] V(\si_{23}) \bigg\} \\
\times &
\exp\left\{-\frac14 \[\right.\ka_{13} \Ga(\si_{13}) + \ka_{23} \Ga(\si_{23})  +  \ka_{12} \Ga(\si_{12})\left.\] \right\} 
\end{aligned}
\end{equation}
Thanks to the symmetry identified earlier, there is essentially only
one integral to do, as the six terms in Eq.~(\ref{e:Beqn}) in are
related by a permutation symmetry.  We can define a function of three
variables
\begin{equation}
\begin{aligned}
 &F(\ka_{12},\ka_{23},\ka_{31})  = \int \ud\si_1 \ud\si_2 \ud\si_3
 \Pi(\si_{13}) V(\si_{12}) \\
 & \times \exp\left\{\frac14 \[-\ka_{31} \Ga(\si_{13}) -\ka_{23}
   \Ga(\si_{23}) -\ka_{12} \Ga(\si_{12})\]\right\}, \label{app1}
\end{aligned}
\end{equation}
in terms of which
\begin{equation}
\begin{aligned}
 & k^A_1k^B_2k^C_3 \lvev{C^{ABC}D} e^{-\half \vev{D^2}} = \frac14 \big\{ \\
 &\ka_{12}\ka_{31}F(\ka_{12},\ka_{23},\ka_{31}) +
 \ka_{12}\ka_{23}F(\ka_{12},\ka_{31},\ka_{23}) \\
 + & \ka_{31}\ka_{12}F(\ka_{31},\ka_{23},\ka_{12}) +
 \ka_{31}\ka_{23}F(\ka_{31},\ka_{12},\ka_{23}) \\
 + & \ka_{23}\ka_{12}F(\ka_{23},\ka_{31},\ka_{12}) +
 \ka_{23}\ka_{31}F(\ka_{23},\ka_{12},\ka_{31}) \big\}.
\end{aligned}
\end{equation}
A change of coordinates simplifies the integration:
\begin{equation}
\si_{123} = \frac13(\si_1+\si_2+\si_3),
\end{equation}
giving
\begin{equation}
\begin{aligned}
& F(\ka_{12},\ka_{23},\ka_{31}) = \int \ud\si_{123} \ud\si_{12}
\ud\si_{31} \Pi(\si_{31}) V(\si_{12}) \\
& \times  \exp\left\{\frac14\[-\ka_{31} \Ga(\si_{31}) -\ka_{23}
 \Ga(\si_{23})  -\ka_{12} \Ga(\si_{12})\] \right\},
\end{aligned}
\end{equation}
where we have used $\Pi(\si_{31})=\Pi(\si_{13})$ and $\Ga(\si_{31})=\Ga(\si_{13})$. 
Inserting the identity 
\begin{equation}
1 = \int \ud\si_{23} \de\(\si_{23}+ \si_{12}+\si_{31} \),
\end{equation}
and using an integral representation for the $\de$-function we have
\begin{equation}
\begin{aligned}
&  F(\ka_{12},\ka_{23},\ka_{31}) = L \int \dfrac{\ud\la}{2\pi}\int
 \ud\si_{12} \ud\si_{23} \ud\si_{31} \\ 
&\times \Pi(\si_{31}) V(\si_{12}) \exp\left[i\la\(\si_{23} + \si_{12} +\si_{31}\)\right] \\
& \times \exp \left\{- \frac14
   \[\ka_{13} \Ga(\si_{31}) + \ka_{23} \Ga(\si_{23}) + \ka_{12}
   \Ga(\si_{12})\]\right\}.
\end{aligned}
\end{equation}
Hence we can factorise
\begin{equation}
\begin{aligned}
 F(\ka_{12},\ka_{23},\ka_{31}) = L \int \dfrac{\ud\la}{2\pi}
 I(\la,{\ka_{12}}) J(\la,{\ka_{23}})K(\la,{\ka_{31}}),
\end{aligned}
\end{equation}
where
\begin{eqnarray}
I(\la,\ka_{12}) & = & \int \ud\si V(\si)e^{{i}\la\si - \frac14 \ka_{12}\Ga(\si)}, \\
J(\la,\ka_{23}) & = & \int \ud\si e^{{i}\la\si - \frac14 \ka_{23}\Ga(\si)},\\
K(\la,\ka_{31}) & = & \int \ud\si \Pi(\si)e^{{i}\la\si - \frac14 \ka_{31}\Ga(\si)}.
\end{eqnarray}
We evaluate $I$, $J$ and $K$ for small angles. For $\ka_{ab}$ large
and positive, and assuming the relevant limits for the correlation
functions
\begin{equation}
 \Ga(\si) \simeq \tb^2\si^2, \quad V(\si) \simeq \vb^2,\quad \Pi(\si)
 \simeq \half  \frac{c_0}{\hat\xi}\si^2, \label{app2}
\end{equation}
one gets
\begin{align}
 I(\la,\ka_{12}) & \simeq \dfrac{\vb^2}{\tb}\sqrt{\dfrac{4
     \pi}{\kappa_{12}}} \exp \left(-
   \dfrac{\lambda^2}{\kappa_{12} \tb^2} \right),\\
 J(\la,\ka_{23}) & \simeq \dfrac{1}{\tb}
 \sqrt{\dfrac{4\pi}{\kappa_{23}}} \exp \left( -
   \dfrac{\lambda^2}{\kappa_{23} \tb^2} \right),\\
 K(\la,\ka_{31}) & \simeq \dfrac{c_0}{\hat{\xi}}\dfrac{1}{\kappa_{31} \tb^3} \sqrt{\dfrac{4
     \pi}{\kappa_{31}}} \left( 1-\frac{2\lambda^2}{\kappa_{31} \tb^2} \right) \exp
 \left(-\dfrac{\lambda^2}{\kappa_{31} \tb^2} \right).
\end{align}
Hence
\begin{equation}
\begin{aligned}
 F(\ka_{12},\ka_{23},\ka_{31}) &\simeq 4\pi L
 \frac{\vb^2}{\tb^4}\frac{c_0}{\hat\xi} \\
& \times \dfrac{\ka_{23}+\ka_{12}}{\(\ka_{23}\ka_{31} +\ka_{12}\ka_{31}+
   \ka_{12}\ka_{23} \)^{3/2}}\,. \label{app3}
\end{aligned}
\end{equation}
Though we have assumed $\kappa_{ab} > 0$ in deriving this equation,
the result is still applicable even if $\kappa_{ab}$ is negative
because the combination $\ka_{23}\ka_{31}
+\ka_{12}\ka_{31}+\ka_{12}\ka_{23}$ in the denominator is positive
definite.  Actually, by using approximations (\ref{app2}), one can
derive (\ref{app3}) directly from Eq.~(\ref{app1}) which is
well-defined even for negative $\kappa_{ab}$ because of the positive
definiteness of the exponent. Defining the new function
\begin{equation}
 g(\ka_{12},\ka_{23},\ka_{31}) = \dfrac{\ka_{12}+\ka_{23}} 
{\(\ka_{23}\ka_{31} +\ka_{12}\ka_{31}+ \ka_{12}\ka_{23} \)^{3/2}},
\label{e:gDef}
\end{equation}
we can write
\begin{equation}
\begin{aligned}
& \BT(\bk_1,\bk_2,\bk_3) = -\epsilon^3 \pi c_0\frac{\vb^2}
{\tb^4}\frac{L\hat\xi}{\Area}\frac{1}{\hat\xi^2} 
\frac{1}{k_1^2k_2^2k_3^2} \\
\times \big\{ 
&\ka_{12}\ka_{31}g(\ka_{12},\ka_{23},\ka_{31}) +
\ka_{12}\ka_{23}g(\ka_{12},\ka_{31},\ka_{23}) \\
+ & \ka_{31}\ka_{12}g(\ka_{31},\ka_{23},\ka_{12}) +
\ka_{31}\ka_{23}g(\ka_{31},\ka_{12},\ka_{23}) \\
+ & \ka_{23}\ka_{12}g(\ka_{23},\ka_{31},\ka_{12}) +
\ka_{23}\ka_{31}g(\ka_{23},\ka_{12},\ka_{31})
\big\}.
\end{aligned}
\end{equation}
Thanks to $g$ being symmetric in its first two arguments, and using
$\ka_{23}+\ka_{12} = k_2^2$ (and circular permutations), some
simplification follows:
\begin{equation}
\begin{aligned}
 &\BT(\bk_1,\bk_2,\bk_3) = -\epsilon^3 \pi
 c_0\frac{\vb^2}{\tb^4}\frac{L\hat\xi}{\Area}\frac{1}{\hat\xi^2}
 \frac{1}{k_1^2k_2^2k_3^2} \times \big\{ \\ &
 k_2^2\ka_{31}g(\ka_{12},\ka_{23},\ka_{31}) +
 k_1^2\ka_{23}g(\ka_{31},\ka_{12},\ka_{23}) \\ + &
 k_3^3\ka_{12}g(\ka_{23},\ka_{31},\ka_{12}) \big\}.
\end{aligned}
\end{equation}
Finally, using the definition of $g$ from Eq.~(\ref{e:gDef}),
\begin{equation}
\begin{aligned}
 \BT(\bk_1,\bk_2,\bk_3) & = -\epsilon^3 \pi
 c_0\frac{\vb^2}{\tb^4}\frac{L\hat\xi}{\Area}\frac{1}{\hat\xi^2}
 \frac{1}{k_1^2k_2^2k_3^2} \\
 & \times \dfrac{k_1^4 \ka_{23} + k_2^4\ka_{31} +
     k_3^4\ka_{12}}{\(\ka_{23}\ka_{31} +\ka_{12}\ka_{31}+
     \ka_{12}\ka_{23} \)^{3/2}} \,.
\end{aligned}
\end{equation}
\bibliography{bibstrings}
\end{document}